\documentclass[11pt,showpacs]{revtex4}
\input epsf.sty

\usepackage{slashed}
\usepackage{ulem}
\usepackage{cancel}
\usepackage{color}

\usepackage{graphicx}
\usepackage{amsfonts}
\usepackage{amssymb}
\usepackage{rotating}
\usepackage{booktabs}
\usepackage{xcolor}
\usepackage{soul}

\def\comment#1{}

\def\tr{\hbox{tr}}

\def\E{{\mathcal E}}

\usepackage{graphicx}
\begin{document}
\title{
An effective strong-coupling theory of composite particles in UV-domain
}
\author{She-Sheng Xue}
\email{xue@icra.it}
\affiliation{ICRANeT, Piazzale della Repubblica, 10-65122, Pescara,\\
Physics Department, Sapienza University of Rome, Piazzale Aldo Moro 5, 00185 Roma, Italy} 


\begin{abstract}
We briefly review 
the effective field theory of massive composite particles, 
their gauge couplings and characteristic energy scale  
in the UV-domain of UV-stable fixed point 
of strong four-fermion coupling,
then mainly focus the discussions 
on the decay channels of composite particles 
into the final states of the SM gauge bosons, leptons and quarks.
We calculate the rates of composite bosons 
decaying into two gauge bosons 
$\gamma\gamma$, $\gamma Z^0$, $W^+W^-$, $Z^0Z^0$ and give the 
ratios of decay rates of different channels 
depending on gauge couplings only. 
It is shown that a composite fermion decays 
into an elementary
fermion and a composite boson, the latter being an intermediate state decays into
two gauge bosons, leading to a peculiar kinematics of final states 
of a quark (or a lepton) and two gauge bosons. 
These provide experimental implications 
of such an effective theory of composite particles beyond the SM. 
We also present some speculative discussions on the 
channels of composite fermions 
decaying into $WW$, $WZ$ and $ZZ$ two boson-tagged jets with quark jets, 
or to four-quark jets. Moreover, at the same energy scale of composite 
particles produced in high-energy experiments,  
composite particles are also produced by high-energy sterile neutrino 
(dark matter) collisions, their decays lead to excesses of cosmic ray particles in space and signals of SM particles 
in underground laboratories.       
\end{abstract}


\maketitle

\newpage

\setcounter{footnote}{0}

\tableofcontents
\newpage

\section{\bf Introduction}
\hskip0.1cm
The parity-violating (chiral) gauge symmetries and spontaneous/explicit 
breaking of these symmetries for the hierarchy of fermion masses have 
been at the center of a conceptual elaboration that has played a major 
role in donating to mankind the beauty of the SM for 
particle physics. The Nambu-Jona-Lasinio (NJL) model \cite{njl} of 
dimension-6 four-fermion operators at high energies and its effective 
counterpart, the phenomenological model \cite{higgs} 
of the elementary Higgs boson and its Yukawa-coupling to fermions 
at low energies, provide an elegant 
and simple description for the electroweak symmetry 
breaking and intermediate gauge boson masses. After a great 
experimental effort for many years, using
data recorded at $\sqrt{s}=7, 8$ TeV experiments of $pp$ collision 
at the Large Hadron Collider (LHC),
the ATLAS \cite{ATLAS} and CMS \cite{CMS} collaborations have 
shown the first observations of a 125 GeV scalar particle in the search 
for the SM Higgs boson. This far-reaching result
begins to shed light on this most elusive and fascinating arena of fundamental particle physics. Recently, in the Run-2 of the upgraded LHC, the preliminary 
results on $\sqrt{s}=13$ TeV $pp$ collision data are prepared 
by ATLAS \cite{ATLAS2016} and CMS \cite{CMS2016} to search for new (beyond the SM) 
resonant and/or 
nonresonant phenomena that might manifest themselves in the high-energy regime with
final states of diboson, dilepton, dijet and so forth. The diphoton channel preliminarily showed the
trace of a new resonance at the diphoton invariant mass 
${\mathcal M}_{\gamma\gamma}$, 
which however has not been
further confirmed by increasing the integrated luminosity and energy 
of $pp$ collisions.

\subsection{Weak-interacting four-fermion operators and symmetry-breaking phase}

The dynamics of new physics at high energies may be represented by an 
effective theory of high-dimensional operators of fermion fields, 
e.g., dimension-6 four-fermion
operators, preserving at least the SM gauge symmetries.  
The strong technicolor dynamics of extended gauge theories at the 
TeV scale was invoked \cite{hill1994,bhl1990a}
to have a natural scheme incorporating the 
relevant four-fermion operator 
$G(\bar\psi^{ia}_Lt_{Ra})(\bar t^b_{R}\psi_{Lib})$ of 
the $\langle\bar t t\rangle$-condensate model \cite{bhl1990}, to generate the top-quark mass via spontaneous symmetry breaking (SSB). 
On the other hand, these relevant operators can be constructed 
on the basis of phenomenology of the SM at low-energies.  
In 1989, several authors \cite{bhl1990,nambu1989,Marciano1989} 
suggested that the SM symmetry breakdown could be 
a dynamical mechanism of the NJL type that 
intimately involves the top quark at the high-energy scale $\Lambda$. 
Since then, many models based on this idea have been 
studied \cite{DSB_review}. In the scenes of renormalization group 
(scaling) invariance and resultant renormalization-group (RG) equations, 
there is a scaling region (IR-domain) 
of the infrared (IR) stable fixed point of four-fermion 
operators, the low-energy SM physics
was supposed to be achieved by the RG 
equations in the IR-domain with the electroweak scale $v\approx 239.5$ 
GeV \cite{bhl1990a,Marciano1989,bhl1990}. It is in this IR-domain, 
we recently present the detailed study of hierarchy mass spectrum of SM fermions: from top quark to electron 
neutrino \cite{xue2016_1,xue2016_2}, on the 
basis of effective four-fermion operators of Einstein-Cartan type, as
will be shown in Sec.~\ref{four-fermionS}.  

\subsection{Strong-interacting four-fermion operators and gauge-symmetric phase}

In addition to the IR-domain of weak-coupling four-fermion operators 
in the SSB phase where the low-energy SM is realized, 
we find a scaling region (UV-domain) of the ultraviolet (UV) stable fixed point of strong-coupling four-fermion operators 
in the gauge-symmetric phase. In this UV-domain at high energies, 
it realizes an effective theory of composite bosons and 
fermions composed by SM elementary fermions, these composite particles 
and their interactions preserving the 
SM gauge symmetries \cite{xue1997,xue2000,xue2000prd,xue2003,xue2014}. This is the issue that we would like to focus in this article.  

In Refs.~\cite{xue2014,xue2015}, we have already presented 
discussions on the possible decay channels of composite bosons 
and fermions into final states of jets and leptons in the 
framework of effective four-fermion operators in high energies. 
In this article, we focus on the discussions on the possible channels of composite bosons decaying into two SM gauge bosons, and 
composite fermions decaying into two SM gauge bosons (diboson) and a quark/lepton,
as well as their experimental implications. 
It is preliminarily shown that the diphoton channel of composite bosons decay should have 
the largest branching ratio, however other diboson channels 
should also contribute to the invariant mass ${\mathcal M}$ of possible composite-boson resonance.  
Instead, composite fermions decay into a composite boson 
and a quark/lepton, the former is an intermediate state, which then decays into two bosons, i.e., a quark-diboson or lepton-diboson channel. These final states and their kinematic 
relations of such a quark-diboson or lepton-diboson channel are peculiar.
This provides a possibility/criterion to verify the effective theory of
composite particles in high-energy experiments.
\comment{ 
so that the effective theory of composite bosons and fermions could be 
possibly verified or excluded, provided that the diboson channels of 
composite bosons decay and the quark-diboson or lepton-diboson 
channel of composite fermions decay can be experimentally 
verified or excluded.
}

\subsection{Article arrangement}

In Sec.~\ref{four-fermionS}, we present 
the SM gauge-symmetric operators of 
four-fermion interactions of Einstein-Cardan type 
and the brief discussion of their possible origin. 
Some explanations and clarifications 
of the low-energy IR-domain for the SM 
and the high-energy UV-domain for an effective field 
theory of composite particles are given in Sec.~\ref{irduv}. After showing the composite particle spectra 
and effective Lagrangian in Secs.~\ref{composite} in the framework of SM fermion content and gauge symmetries, 
we discuss in Secs.~\ref{cbd} and \ref{cfd} the main results that the composite boson and fermion decay 
into SM gauge bosons and/or elementary fermions
in connection with high-energy 
experiments.
In the final section, we present some discussions and speculations on 
the possible channels of composite particles interacting with SM particles,
that could be relevant to some experiments at the energy scale of composite particles.  

\noindent
\section {Four-fermion operators beyond the SM 
}\label{four-fermionS}

In order for a self-contained and self-consistent article, as well as for readers' convenience, we include this section of describing effective 
four-fermion operators at the cutoff $\Lambda$ that was similarly presented in Ref.~\cite{xue2016_2} for discussing the hierarchy 
spectrum of SM fermions in the IR-domain. It is also necessary to present 
these effective four-fermion operators at the cutoff $\Lambda$ 
to clarify its induced Lagrangian of the composite particle spectrum and interactions in the UV-domain.
    
\subsection{Regularization and quantum gravity}\label{qgravity}

Up to now the theoretical and experimental studies tell us 
the chiral gauge-field interactions to fermions 
in the lepton-quark family that is replicated three times and mixed. 
The spontaneous breaking of these chiral gauge symmetries and generating of fermion masses are made by the Higgs field sector. 
In the IR-fixed-point domain of weak four-fermion coupling 
or equivalently weak Yukawa coupling, the SM Lagrangian with all relevant operators (parametrizations) is realized and behaves an effective and renormalizable field theory in low energies. 
To achieve these SM relevant operators, a finite field theory of chiral-gauge interactions should be well-defined by including the quantum gravity that naturally provides a space-time regularization (UV cutoff).
As an example, the finite superstring theory is proposed by postulating that instead of a simple space-time point, the fundamental space-time ``constituents" is a space-time ``string". 
The Planck scale is a plausible cut-off, at which 
all principle and symmetries are fully respected by gauge fields and 
particle spectra, fermions and bosons. \comment{NPB The superstring compactification
on the Calabi-Yau manifold or orbifold results in the pointlike supergravity theories, inducing 
high-dimensional operators below the
Planck scale that provides a perturbatively consistent scheme to understand the SM parameters, e.g.~particle masses, and their logarithmic running with energy, see review Ref.~\cite{faraggi2014}}

In this article, we do not discuss how a fundamental theory
at the Planck scale 
induces high-dimensional operators. Instead, as a postulation or motivation, we argue the presence of at least four-fermion operators beyond the SM from the following point view. A well-defined quantum field theory for the SM Lagrangian requires a natural regularization (UV cutoff $\Lambda$) fully preserving the SM 
chiral-gauge symmetry. The quantum gravity naturally provides a such regularization of discrete 
space-time with the minimal length $\tilde a\approx 1.2\,a_{\rm pl}$ 
\cite{xue2010}, 
where the Planck length 
$a_{\rm pl}\sim 10^{-33}\,$cm and scale $\Lambda_{\rm pl}=\pi/a_{\rm pl}\sim 10^{19}\,$GeV. 
However, the no-go theorem \cite{nn1981} tells us 
that there is no any consistent way to regularize 
the SM bilinear fermion Lagrangian to exactly preserve the SM chiral-gauge symmetries, which 
must be explicitly broken at the scale of fundamental space-time cutoff $\tilde a$. 
This implies that the natural quantum-gravity regularization for the SM should
lead us to consider at least dimension-6 four-fermion operators originated from quantum gravity effects at short distances \footnote{In the regularized and quantized EC theory \cite{xue2010}
with a basic space-time cutoff, in addition to dimension-6 four-fermion operators,
there are high-dimensional fermion operators ($d>6$), 
e.g., $\partial_\sigma J^\mu\partial^\sigma J{_\mu}$, which are suppressed
at least by ${\mathcal O}(\tilde a^4)$.
}. As a model, we adopt the four-fermion operators of the torsion-free Einstein-Cartan Lagrangian within the framework of the SM fermion content and gauge symmetries. We stress that a fundamental theory at the UV cutoff is still unknown.
\comment{ 
The
quantum-gravity origin of four-fermion operators is just an argumentation or a speculation. This
however motivates us to adopt the EC Lagrangian as an effective Lagrangian with SM chiral
gauge symmetries and fermion content to study the SM fermion masses and their hierarchy.
This article does not focus on the origin of the four-fermion operators from the underlying
theory, nevertheless, it is expected that the most promising candidate for the fundamental theory
of quantum gravity should be the superstring theory, and there have been many detailed studies of
the origin of fermion generations and masses in the context of superstring theory and the resultant
pointlike supergravity theories, see for example the recent monograph [77]on the issue. 
}

\subsection{Einstein-Cartan theory with SM gauge symmetries and fermion content}

The Lagrangian of torsion-free Einstein-Cartan (EC) theory reads,
\comment{The four-fermion operators of the classical and 
torsion-free Einstein-Cartan (EC) theory are naturally
obtained by integrating over ``static'' torsion fields at the Planck length.}
\begin{eqnarray}
{\mathcal L}_{EC}(e,\omega,\psi)&=&  {\mathcal L}_{EC}(e,\omega) + 
\bar\psi e^\mu {\mathcal D}_\mu\psi +GJ^dJ_d,
\label{ec0}
\end{eqnarray}
where the gravitational Lagrangian ${\mathcal L}_{EC}={\mathcal L}_{EC}(e,\omega)$, 
tetrad field $e_\mu (x)= e_\mu^{\,\,\,a}(x)\gamma_a$,
spin-connection field $\omega_\mu(x) = \omega^{ab}_\mu(x)\sigma_{ab}$,  
the covariant derivative ${\mathcal D}_\mu =\partial_\mu - ig\omega_\mu$ and 
the axial current $J^d=\bar\psi\gamma^d\gamma^5\psi$ of massless fermion fields. 
The four-fermion coupling $G$ relates to the gravitation-fermion gauge 
coupling $g$ and fundamental space-time cutoff $\tilde a$. 

In the context of the SM one-family fermion content and gauge symmetries, 
we consider massless, two-component, left- and right-handed Weyl 
fermions $\psi^f_{_L}$ 
(doublets) and $\psi^f_{_R}$ (singlets) 
carrying the quantum numbers of the SM $SU_L(2)\times U_Y(1)$ chiral
gauge symmetries, where ``$f$'' is the fermion-family index, as well as three right-handed Weyl sterile neutrinos 
$\nu^f_{_R}$ and their left-handed conjugated fields 
$\nu^{f\, c}_{_R}=i\gamma_2(\nu_{_R})^*$, which do not carry any quantum number of SM gauge symmetries. 
Analogously to the EC theory (\ref{ec0}), we obtain a torsion-free, 
diffeomorphism and {\it local} gauge-invariant 
Lagrangian 
\begin{eqnarray}
{\mathcal L}
&=&{\mathcal L}_{EC}(e,\omega)+\sum_f\bar\psi^f_{_{L,R}} e^\mu {\mathcal D}_\mu\psi^f_{_{L,R}} 
+ \sum_f\bar\nu^{f c}_{_{R}} e^\mu {\mathcal D}_\mu\nu^{f c}_{_{R}}\nonumber\\
&+&G\left(J^{\mu}_{_{L}}J_{_{L,\mu}} + J^{\mu}_{_{R}}J_{_{R,\mu}} 
+ 2 J^{\mu}_{_{L}}J_{_{R,\mu}}\right)\nonumber\\
&+&G\left(j^{\mu}_{_{L}}j_{_{L,\mu}} + 2J^{\mu}_{_L}j_{_{L,\mu}} 
+ 2 J^{\mu}_{_R}j_{_{L,\mu}}\right),
\label{art}
\end{eqnarray}
where the SM gauge fields $A_\mu$ are present 
in the co-variant derivative ${\mathcal D}_\mu$
to preserve the SM gauge symmetries,
and axial currents read
\begin{eqnarray}
J^{\mu}_{_{L,R}}\equiv \sum_f\bar\psi^f_{_{L,R}}\gamma^\mu\gamma^5\psi^f_{_{L,R}}, \quad
j^{\mu}_{_L}\equiv \sum_f\bar\nu^{fc}_{_R}\gamma^\mu\gamma^5\nu^{fc}_{_R}. 
\label{acur}
\end{eqnarray}
The four-fermion coupling $G$ is unique for all four-fermion operators and 
high-dimensional fermion operators ($d>6$) are neglected. 
\comment{If torsion fields that couple to fermion fields 
are not exactly static, propagating a short distance 
$\tilde \ell \gtrsim \tilde a$, 
characterized by their large masses 
$\Lambda\propto \tilde \ell^{-1}$, this implies the four-fermion 
coupling $G\propto \Lambda^{-2}$.
We will in future 
address the issue how the space-time cutoff $\tilde a$ due to quantum 
gravity relates to the cutoff scale $\Lambda(\tilde a)$ 
possibly by intermediate torsion fields 
or the Wilson-Kadanoff renormalization group approach.}

By using the Fierz theorem \cite{itzykson_k,itzykson}, the dimension-6 four-fermion operators in Eq.~(\ref{art}) can be written as \cite{xue2015} 
\begin{eqnarray}
&+&(G/2)\left(J^{\mu}_{_{L}}J_{_{L,\mu}} + J^{\mu}_{_{R}}J_{_{R,\mu}} 
+ j^{\mu}_{_{L}}j_{_{L,\mu}} + 2J^{\mu}_{_L}j_{_{L,\mu}}\right)\label{art0}\\
&-&G\sum_{ff'}\left(\, \bar\psi^f_{_L}\psi^{f'}_{_R}\bar\psi^{f'}_{_R} \psi^f_{_L}
+\, \bar\nu^{fc}_{_R}\psi^{f'}_{_R}\bar\psi^{f'}_{_R} \nu^{fc}_{_R}\right),
\label{art0'}
\end{eqnarray}
which preserve the SM gauge symmetries. Equations (\ref{art0}) and 
(\ref{art0'}) represent repulsive and attractive operators respectively.
The former (\ref{art0}) 
are suppressed by the cutoff ${\mathcal O}(\Lambda^{-2})$, and cannot become relevant and renormalizable operators of effective dimension-4. 
\comment{and in the UV-domain where the 
formation of composite fermions occurs. Therefore they are irrelevant operators. 
Instead, four-fermion operators (\ref{art0'}) are
relevant operators in the both domains of IR- and UV-stable fixed points, respectively 
associated with the SSB dynamics and formation of composite fermions. 
We presented some discussions of these relevant operators and their 
resonant and nonresonant new phenomena for experimental searches. 
We will proceed further discussions in the last section 
of this article. However, main attention will be given to the mass generation of the 
third fermion family in the IR-domain where the SSB dynamics occurs.
}
Thus the torsion-free EC theory with the attractive
four-fermion operators reads,
\begin{eqnarray}
{\mathcal L}
&=&{\mathcal L}_{EC}+\sum_{f}\bar\psi^f_{_{L,R}} e^\mu {\mathcal D}_\mu\psi^f_{_{L,R}} 
+ \sum_{f}\bar\nu^{ fc}_{_{R}} e^\mu {\mathcal D}_\mu\nu^{ fc}_{_{R}}\nonumber\\
&-&G\sum_{ff'}\left(\, \bar\psi^{f}_{_L}\psi^{f'}_{_R}\bar\psi^{f'}_{_R} \psi^{f}_{_L}
+\, \bar\nu^{fc}_{_R}\psi^{f'}_{_R}\bar\psi^{f'}_{_R} \nu^{fc}_{_R}\right)+{\rm h.c.},
\label{art1}
\end{eqnarray}
where the two component Weyl fermions $\psi^{f}_{_L}$ and $\psi^{f}_{_R}$  
respectively are the $SU_L(2)\times U_Y(1)$ gauged doublets and singlets of the SM. 
For the sake of compact notations, $\psi^{f}_{_R}$ are also used to represent 
$\nu^f_R$, 
which have no any SM quantum numbers. 
All fermions are massless, 
they are four-component Dirac fermions 
$\psi^f=(\psi_L^f+\psi_R^f)$, two-component left-handed Weyl neutrinos $\nu^f_L$ 
and four-component sterile Majorana neutrinos $\nu_M^f=(\nu_R^{fc}+\nu_R^f)$ whose 
kinetic terms read
\begin{eqnarray}
\bar\nu^f_{_{L}} e^\mu {\mathcal D}_\mu\nu^f_{_{L}}, \quad
\bar\nu_{_{M}}^f e^\mu {\mathcal D}_\mu\nu_{_{M}}^f =
\bar\nu^f_{_{R}} e^\mu {\mathcal D}_\mu\nu^f_{_{R}} 
+ \bar\nu^{ fc}_{_{R}} e^\mu {\mathcal D}_\mu\nu^{ fc}_{_{R}}.
\label{mnu}
\end{eqnarray} 
In Eq.~(\ref{art1}), $f$ and $f'$ ($f,f'=1,2,3$) are 
fermion-family indexes summed over respectively for three 
lepton families (charge $q=0,-1$) and three quark families ($q=2/3,-1/3$). 
Equation (\ref{art1}) preserves not only the SM gauge symmetries 
and global fermion-family symmetries, but also the global symmetries for 
fermion-number conservations. We adopt the effective four-fermion operators (\ref{art1}) 
in the context of a well-defined quantum field theory 
at the high-energy scale $\Lambda$.
\comment{Relating to the gravitation-fermion gauge 
coupling $g$, the effective four-fermion coupling $G$ is unique 
for all four-fermion operators, and its strength depends on energy scale and 
characterizes: (i) the domain of IR fixed point where the 
spontaneous breaking of SM gauge-symmetries occurs (see for example 
\cite{bhl1990}) 
and (ii) the domain of UV fixed point where the SM gauge-symmetries 
are restored and massive (TeV) composite Dirac fermions are formed \cite{xue2014}. 
}

\subsection{Fermion-family symmetry and mass eigenstate}\label{fsymmetry}

As argued in the introduction Sec.~\ref{qgravity}, the origin of effective 
four-fermion operators in Eqs.~(\ref{ec0}-\ref{art1}) is due to the quantum gravity that couples to all fermion fields and 
provides a natural regularization for chiral gauge field theories, 
like the SM, at the UV cutoff $\Lambda$. Therefore, there is  
no any reason to assume different four-fermion coupling $G$'s for 
different fermions that equally couple to the gravitational field in the torson-free Einstein-Cartan theory (\ref{ec0}). 

It should be further clarified that in the effective Lagrangian 
(\ref{art1}) at the cutoff $\Lambda$, (i) the massless SM 
fermion fields are interacting {\it gauge eigenstates} of the SM gauge symmetries $SU_c(3)\times SU_L(2)\times U_Y(1)$; (ii) due to the unique four-fermion coupling $G$, there are exact global 
fermion-family $U_L(3)\times U_R(3)$ chiral symmetries, i.e., 
flavor or horizon symmetries with respect to different 
charges $q=0,-1,2/3,-1/3$ of fermions in different families. 

In order to study the gauge-boson-fermion and four-fermion interactions in terms of fermion mass-energy spectra and currents, measured 
as physical final states, we adopt the energy-{\it mass eigenstates} 
of fermions in the following discussions of entire article. 
The unitary chiral transformations 
${\mathcal U}_L\in U_L(3)$ and 
${\mathcal U}_R\in U_R(3)$, where ${\mathcal U}_L$ and ${\mathcal U}_R$ 
are related by a unitary matrix ${\mathcal V}$, can be performed 
from gauge eigenstates to mass eigenstates 
(up- and down-quark sectors as example): 
\begin{eqnarray}
\psi^u_L&\rightarrow & {\mathcal U}^{u}_L~\psi^u_L,\quad 
\psi^u_R\rightarrow  {\mathcal U}^{u}_R~\psi^u_R;\quad 
{\mathcal U}^{u}_{L,R}\in U^{u}_{L,R}(3),
\label{cqu}
\end{eqnarray}
and
\begin{eqnarray}
\psi^d_L&\rightarrow & {\mathcal U}^{d}_L~\psi^d_L,\quad 
\psi^d_R\rightarrow  {\mathcal U}^{d}_R~\psi^d_R;\quad 
{\mathcal U}^{d}_{L,R}\in U^{d}_{L,R}(3),
\label{cqd}
\end{eqnarray}
so that in Eq.~(\ref{art1}) the fermion-family indexes $f=f'$, i.e., $\delta_{ff'}$ 
respectively for the $u$-quark sector and 
the $d$-quark sector. As a result, all quark fields are mass eigenstates, 
the four-fermion operators (\ref{art1}) are ``diagonal'' only for each quark family without fermion-family mixing, 
\begin{eqnarray}
-G\, \sum_{f=1,2,3}\Big[\bar\psi^{f}_{_L}\psi^{f}_{_R}\bar\psi^{f}_{_R} \psi^{f}_{_L}\Big]^{\rm up}_{2/3}
-G\, \sum_{f=1,2,3}\Big[\bar\psi^{f}_{_L}\psi^{f}_{_R}\bar\psi^{f}_{_R} \psi^{f}_{_L}\Big]^{\rm down}_{-1/3}.
\label{q1}
\end{eqnarray}
In the following, we adopt this representation and our notations 
will be only for the first SM family, however, they are the same 
for the second and third families.

In this representation, bilinear fermion-mass operators 
$\langle\psi^{f}_{_R} \bar \psi^{f}_{_L}\rangle+{\rm h.c.}$, i.e., quark-mass matrices are diagonalized in the fermion-family space by the biunitary transformations
\begin{eqnarray}
M^u\Rightarrow M^u_{\rm diag}= (m^u_1, m^c_2,m^t_3)={\mathcal U}^{u\dagger}_L M^u {\mathcal U}^u_R,\label{mqu}\\
M^d\Rightarrow M^d_{\rm diag}=(m^d_1, m^s_2,m^b_3)= {\mathcal U}^{d\dagger}_L M^d {\mathcal U}^d_R,\label{mqd}
\end{eqnarray}
where all quark masses (eigenvalues) are positive, ${\mathcal U}_L$ 
and ${\mathcal U}_R$ are related by
\begin{eqnarray}
{\mathcal U}^{u,d}_L={\mathcal V}^{u,d} {\mathcal U}^{u,d}_R,
\label{ulur}
\end{eqnarray}
and ${\mathcal V}^{u,d}$ is a unitary matrix, 
see for example \cite{zli,mohapatra}.
Using unitary matrices ${\mathcal U}^u_{L,R}$ (\ref{cqu}) 
and ${\mathcal U}^d_{L,R}$ (\ref{cqd}), up to a diagonal phase matrix we define the unitary quark-family mixing matrices,
\begin{eqnarray}
{\mathcal U}^{u\dagger}_L{\mathcal U}^{d}_L, &\quad&  
{\mathcal U}^{u\dagger}_L{\mathcal U}^d_R,\nonumber\\
{\mathcal U}^{u\dagger}_R{\mathcal U}^d_L, &\quad&
{\mathcal U}^{u\dagger}_R{\mathcal U}^{d}_R.
\label{mmud}
\end{eqnarray}
where the first element is the CKM matrix $U=U^q_L\equiv 
{\mathcal U}_L^{u\dagger}{\mathcal U}_L^d$. The similar discussions 
for the lepton sector can be found in Ref.~\cite{xue2016_2}.

In the IR effective theory in the IR-domain, the nonzero and different expectational values of fermion-mass operators $\langle\psi^{f}_{_R} \bar \psi^{f}_{_L}\rangle\propto m^f\not=0$ 
(\ref{mqu}) and fermion-family hierarchy $m^f\not=m^{f'}$ 
can be developed due to both spontaneous-symmetry and 
explicit-symmetry breaking of chiral symmetries \cite{xue2016_1}. 
In this case, apart from the breaking of the SM chiral gauge 
symmetries, fermion-family (flavor) $U_L(3)\times U_R(3)$ 
symmetries are broken to the $U(1)$ symmetry for each fermion family, i.e., $U_1(1)\times U_2(1)\times U_3(1)$ for the four-fermion operators 
(\ref{q1}), two-fermion operators (\ref{mqu}) and (\ref{mqd}). 
However, the mass eigenstates of the SM {\it elementary} fermions are different from their chiral-gauge eigenstates, i.e., the interaction vertexes of the chiral-gauge-boson $W^\pm$ and massive fermions are not diagonal in the fermion-family space based on the mass eigenstates of fermions, 
as can be seen in the kinetic terms of 
the effective Lagrangian (\ref{art1}).

In the UV effective theory in the UV-domain, instead, the chiral and flavor symmetries are preserved  for $m^f=0$, indicating that gauge and mass eigenstates 
of {\it composite} particles are the same.   
In Secs.~(\ref{irfcnc}) and (\ref{uvfcnc}), we will come back to 
the discussions of the 
flavor-changing-neutral-current (FCNC) processes, considering 
the chiral- and flavor-symmetries breaking (preserving)
of IR (UV) effective theory in the IR (UV) domain respectively.
   
\noindent
\subsection{SM gauge-symmetric four-fermion operators}\label{sm4f}
\hskip0.1cm
Using Eq.~(\ref{q1}) and 
we explicitly show SM gauge symmetric four-fermion operators.
In the quark sector, the four-fermion operators are  
\begin{eqnarray}
G\left[(\bar\psi^{ia}_Lt_{Ra})(\bar t^b_{R}\psi_{Lib})
+ (\bar\psi^{ia}_Lb_{Ra})(\bar b^b_{R}\psi_{Lib})\right]+{\rm ``terms"},
\label{bhlx}
\end{eqnarray}
where $a,b$ and $i,j$ are the color and flavor indexes 
of the top and bottom quarks, the quark $SU_L(2)$ doublet 
$\psi^{ia}_L=(t^{a}_L,b^{a}_L)$ 
and singlet $\psi^{a}_R=t^{a}_R,b^{a}_R$ are the eigenstates 
of electroweak interaction. 
The first and second terms in Eq.~(\ref{bhlx}) are respectively 
the four-fermion operators of top-quark channel \cite{bhl1990} 
and bottom-quark channel,
whereas ``terms" stands for 
the first and second quark families that can be obtained 
by substituting $t\rightarrow u,c$ 
and $b\rightarrow d,s$ \cite{xue2013_1,xue2013,xue2014}. 

In the lepton sector with three right-handed sterile neutrinos $\nu^\ell_R$ 
($\ell=e,\mu,\tau$), the four-fermion operators 
in terms of gauge eigenstates are,
\begin{eqnarray}
G\left[(\bar\ell^{i}_L\ell_{R})(\bar \ell_{R}\ell_{Li})
+ (\bar\ell^{i}_L\nu^\ell_{R})(\bar \nu^\ell_{R}\ell_{Li}) 
+ (\bar\nu^{\ell\, c}_R\nu^{\ell\,}_{R})(\bar \nu^{\ell}_{R}\nu^{\ell c}_{R})\right],
\label{bhlxl}
\end{eqnarray}
preserving all SM gauge symmetries, 
where the lepton $SU_L(2)$ doublets $\ell^i_L=(\nu^\ell_L,\ell_L)$, singlets 
$\ell_{R}$ and the conjugate fields of sterile neutrinos 
$\nu_R^{\ell c}=i\gamma_2(\nu_R^{\ell})^*$.  
Coming from the second term in Eq.~(\ref{art1}), the last term in Eq.~(\ref{bhlxl}) 
preserves the symmetry 
$U_{\rm lepton}(1)$ for the lepton-number 
conservation, although $(\bar \nu^{\ell}_{R}\nu^{\ell c}_{R})$ 
violates the lepton number of family ``$\ell$'' by two units. 

Similarly, from the second term in Eq.~(\ref{art1}) there are following four-fermion operators  
\begin{eqnarray}
G\left[(\bar\nu^{\ell\, c}_R\ell_{R})(\bar \ell_{R}\nu^{\ell\, c}_{R})
+(\bar\nu^{\ell\, c}_R u^{\ell}_{a,R})(\bar u^{\ell}_{a,R}\nu^{\ell c}_{R})
+(\bar\nu^{\ell\,c}_R d^\ell_{a,R})(\bar d^{\ell}_{a,R}\nu^{\ell c}_{R})\right],
\label{bhlbv}
\end{eqnarray}
where quark fields $u^{\ell}_{a,R}=(u,c,t)_{a,R}$ and $d^{\ell}_{a,R}=(d,s,b)_{a,R}$.

\subsection{Four-fermion operators of quark-lepton interactions}

Although the four-fermion operators in Eq.~(\ref{art1}) do not have 
quark-lepton interactions, we consider the following
SM gauge-symmetric four-fermion 
operators that contain quark-lepton 
interactions \cite{xue1999nu,xue2016_1,xue2016_2}, 
\begin{eqnarray}
G\left[(\bar\ell^{i}_Le_{R})(\bar d^a_{R}\psi_{Lia})
+(\bar\ell^{i}_L\nu^e_{R})(\bar u^a_{R}\psi_{Lia})\right] +(\cdot\cdot\cdot),
\label{bhlql}
\end{eqnarray}
where $\ell^i_L=(\nu^e_L,e_L)$ and $\psi_{Lia}=(u_{La},d_{La})$
for the first family. The ($\cdot\cdot\cdot$) represents   
for the second and third families with substitutions: 
$e\rightarrow \mu, \tau$, $\nu^e\rightarrow \nu^\mu, \nu^\tau$, and 
$u\rightarrow c, t$ and $d\rightarrow s, b$. 
The four-fermion operators (\ref{bhlql}) of quark-lepton interactions 
are not included in Eq.~(\ref{art1}), since leptons and quarks are in separated
representations of SM gauge groups. They should be expected
in the framework of Einstein-Cartan theory and $SO(10)$ unification theory \cite{lq1997}.

\comment{By using unitary chiral transformations, the quark-lepton interacting operators (\ref{bhlql}) for each fermion family can be generalized to 
\begin{eqnarray}
G\sum_{ff'}\Big\{(\bar\ell^{if}_Le^{f'}_{R})(\bar d^{af'}_{R}\psi^f_{Lia})
+(\bar\ell^{if}_L\nu^{f'}_{eR})(\bar u^{af'}_{R}\psi^f_{Lia})\Big\},
\label{bhlqlm}
\end{eqnarray}
analogously to the four-fermion operators in Eq.~(\ref{art1}). For detailed discussions on the CKM-like mixing matrix analogous to Eq.~(\ref{mmud}), please see the recent article \cite{xue2016_2}.
}

\section{\bf IR-stable and UV-stable fixed points and their scaling regions}\label{irduv}

Apart from what is possible new physics at the UV scale $\Lambda$ explaining the 
origin of these effective four-fermion operators, 
it is essential and necessary to study: (i) the phase diagram in the space of 
these effective four-fermion operator couplings; (ii) which dynamics of these operators 
undergo in terms of their couplings as functions of running energy 
scale $\mu$; (iii) associating to these dynamics where infrared (IR) 
and/or ultraviolet (UV) stable fixed point of physical couplings locates; 
(iv) in the IR and/or UV domains (scaling regions) of these stable fixed points, 
which operators become physically relevant and renormalizable following RG equations
(scaling laws), 
and other irrelevant operators are suppressed by the cutoff at least 
$\mathcal O(\Lambda^{-2})$.  

\subsection{ Symmetry-breaking phase and the IR-domain of the IR-stable fixed point}\label{ird}

In the NJL symmetry-breaking phase of weak coupling $G>G_c$ \footnote{In
the symmetric phase of extremal weak coupling $G<G_c$, all 
four-fermion interacting amplitudes are suppressed by the cutoff 
$\Lambda$.}, 
where $G_c$ is the weak critical coupling of the NJL dynamics, 
the low-energy SM physics
was supposed to be achieved in the IR-domain ($G\gtrsim G_c$) 
of IR-stable fixed point \cite{bhl1990a,Marciano1989,bhl1990}.
Bardeen, Hill and Lindner (BHL) proposed the effective Lagrangian of the $t\bar t$-condensate model \cite{bhl1990},
\begin{eqnarray}
L &=& L_{\rm kinetic} + g_{t0}(\bar \psi_L t_RH+ {\rm h.c.})
+\Delta L_{\rm gauge}\nonumber\\ 
&+& Z_H|D_\mu H|^2-m_{_0H}^2H^\dagger H
-\frac{\lambda_0}{2}(H^\dagger H)^2,
\label{bhl}
\end{eqnarray} 
with a massive composite Higgs boson $H=(\bar t t)$ and massless 
Goldstone bosons $(\bar t\gamma_5t)$ and $(\bar t\gamma_5b)$, 
which are the longitudinal modes of massive intermediate gauge bosons. 
The composite Higgs' bare form-factor (wave-function renormalization) 
$Z_H$ and masss 
$m_{_0H}$, as well as its Yukawa $g_{t0}$ and quartic $\lambda_0$ couplings are defined at an intermediate scale $\E$ ($v<\E<\Lambda$), 
that will be clear below (\ref{thvari}). 
Renormalized quantities, like Yukawa coupling $\bar g_t(\mu^2)$ 
and quartic couplings $\bar \lambda(\mu^2)$
are defined at the running energy scale $\mu$ and satisfy the RG equations 
in the IR-domain,
\begin{eqnarray}
16\pi^2\frac{d\bar g_t}{dt} &=&\left(\frac{9}{2}\bar g_t^2-8 \bar g^2_3 - \frac{9}{4}\bar g^2_2 -\frac{17}{12}\bar g^2_1 \right)\bar g_t,
\label{reg1}\\
16\pi^2\frac{d\bar \lambda}{dt} &=&12\left[\bar\lambda^2+(\bar g_t^2-A)\bar\lambda + B -\bar g^4_t \right],\quad t=\ln\mu \label{reg2}
\end{eqnarray}
where one can find $A$, $B$ and RG equations for 
running gauge couplings $g^2_{1,2,3}$ in Eqs.~(4.7), (4.8) of 
Ref.~\cite{bhl1990}. 

This IR-domain of IR-stable fixed point $G_c$ 
can be illustrated by a positive $\beta(G)$-function 
for $G\gtrsim G_c$. Taking into account 
one fermion-loop contribution (tadpole diagram) 
from the SSB-relevant top-quark channel (\ref{bhlx}), we calculate
the gap equation  for top-quark mass $m_t\not=0$, 
and obtain the $\beta$-function 
\begin{eqnarray}
\beta(G)\equiv \mu \frac{dG}{d\mu} 
&\approx & 2\frac{G^2}{G_c}\left(\frac{\mu}{\E}\right)^2
\left[1+\ln \left(\frac{\E}{\mu}\right)^2\right]>0,
\label{gbeta}
\end{eqnarray}
for $ G \gtrsim G_c\equiv 8\pi^2/(N_c\E^2)$ and intermediate energy 
scale $\E > \mu \gtrsim v$, where $N_c=3$ is the number of colors.
The positive $\beta$-function of Eq.~(\ref{gbeta}) 
indicates that an IR-stable fixed point $G_c$ and the IR-domain 
$G\rightarrow G_c+0^+$ as the running energy scale 
$\mu\rightarrow v$, as shown the part 
``I'' of the $\beta(G)$-function in Fig.~\ref{fixp}. As the energy scale
$\mu$ decreases, indicated by a thick arrow in Fig.~\ref{fixp}, 
the RG flow is attracted to the IR-stable fixed point, and the effective SM of particle physics in low energies 
is realized in this IR-domain (vicinity) of IR-stable fixed point.

\subsubsection{Resolution to the $t\bar t$-condensate model}

Both experimental $m_t$ and $m_{_H}$ values of top-quark and Higgs-boson 
masses were unknown in the early 1990s. In order to find low-energy values 
$m_t$ and $m_{_H}$ close to the IR-stable fixed point, 
BHL \cite{bhl1990} imposed the 
compositeness conditions 
$\tilde Z_{H}(\mu)=1/\bar g_t^2(\mu)\rightarrow 0$ 
and $\tilde \lambda(\mu)\equiv \bar\lambda(\mu)/\bar g^4_t(\mu) \rightarrow 0$, as energy scale $\mu\rightarrow\Lambda$ 
for different values of
the high-energy cutoff $\Lambda$ as the boundary conditions to 
solve the RG equations (\ref{reg1}) and (\ref{reg2}). 
As a result, masses $m_t, m_{_H}$ and Higgs-decay 
width obtained do not agree with experimental values, thus the BHL 
$t\bar t$-condensate model was firmly excluded. From the theoretical point of view, the compositeness conditions implies that the BHL composite Higgs boson is a {\it loosely} bound state of 
top and anti-top-quark pair ($\bar t t$) because its wave-function
renormalization $\tilde Z_{H}(\mu)$ goes to zero as $\mu\rightarrow \Lambda$.

\begin{figure}
\begin{center}
\includegraphics[height=1.40in]{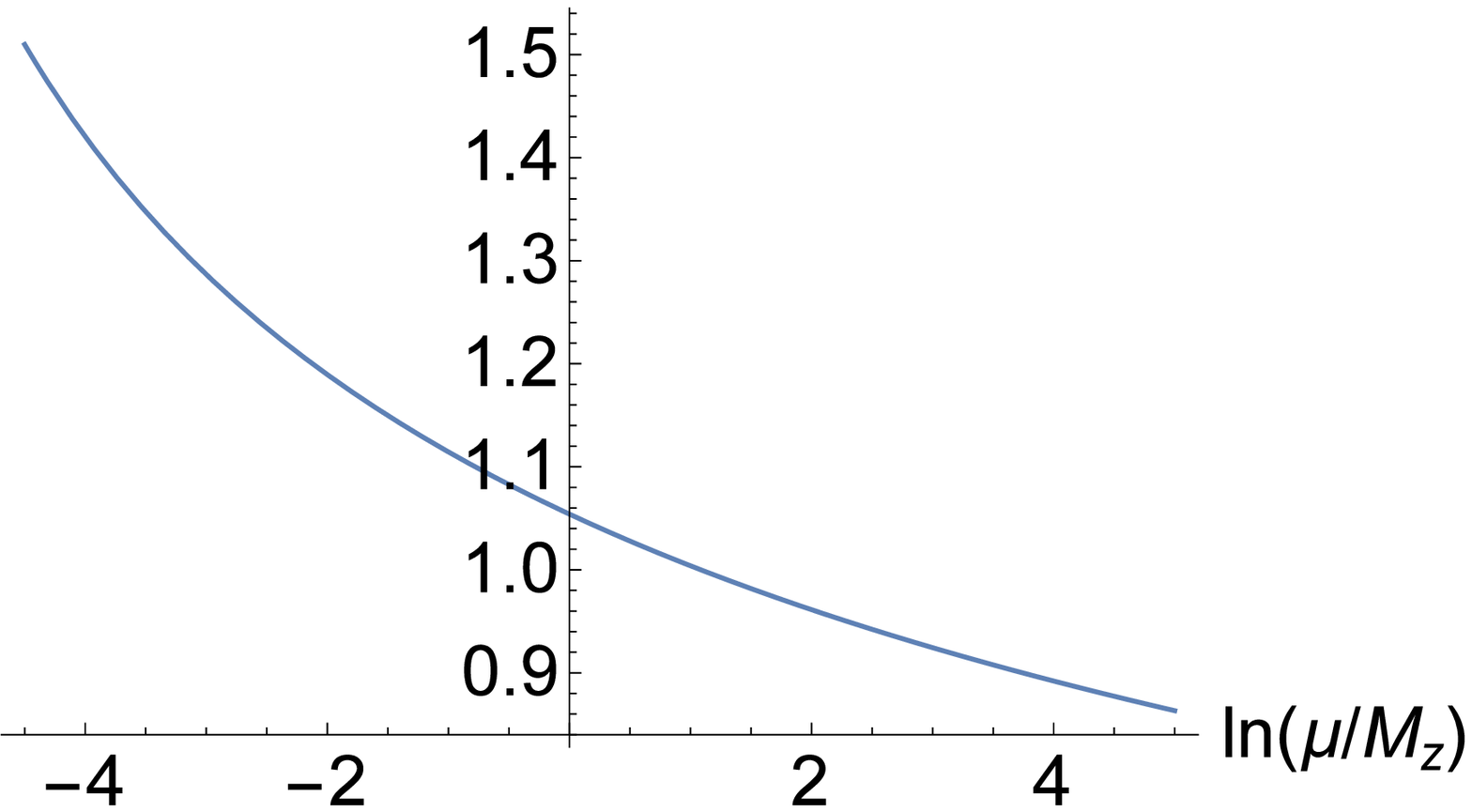}
\includegraphics[height=1.40in]{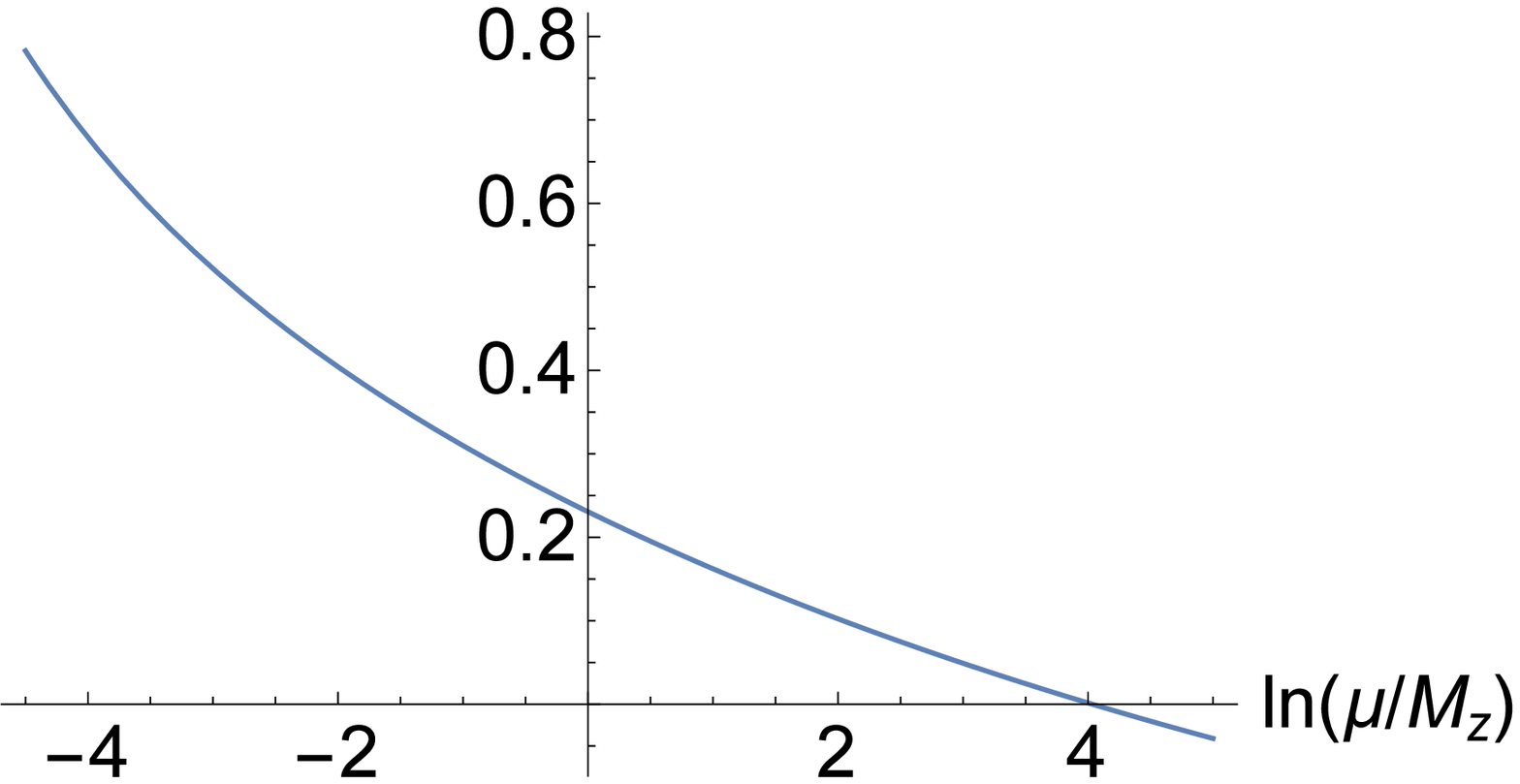}
\put(-120,105){\footnotesize $\tilde\lambda(\mu)$}
\put(-320,105){\footnotesize $\bar g_t(\mu)$}
\caption{Using
experimentally measured SM quantities (including $m_t$ and $m_{_H}$) 
as boundary values, we uniquely solve the RG equations
for the composite Higgs-boson model \cite{bhl1990}, 
we find \cite{xue2013,xue2014} the effective 
top-quark Yukawa coupling $\bar g_t(\mu)$ (left) 
and effective Higgs quartic coupling $\tilde \lambda(\mu)$ (right) in the range
$1.0 ~{\rm GeV}\lesssim \mu \lesssim 13.5~ {\rm TeV}$.
Note that $\tilde\lambda(\E)=0$ at $\E\approx 5.14~{\rm TeV}$ and $\tilde\lambda(\mu)<0$ 
for $\mu >\E$. 
} \label{figyt}
\end{center}
\end{figure}

Instead, we adopted \cite{xue2013, xue2014} the experimental values of
the top-quark and Higgs-boson masses, 
\begin{eqnarray}
m_{_H}=126\pm 0.5 \,{\rm GeV};\quad m_t=172.9\pm 0.8\, {\rm GeV},
\label{varmass}
\end{eqnarray}
and used the mass-shell conditions
$m_t=\bar g_t(m_t)v/\sqrt{2}$ and $m_{_H}^2/2=\tilde \lambda (m_{_H}) v^2,
$ as infrared boundary conditions to integrate the RG equations 
(\ref{reg1}) and (\ref{reg2}) so as to uniquely determine 
the functions of top-quark Yukawa $\bar g_t(\mu)$ 
and Higgs-quartic $\tilde\lambda(\mu)$ couplings (see Fig.~\ref{figyt}), 
as well as to obtain the finite values $\tilde Z_{H}(\E)\approx 1.26$ 
and the intermediate energy scale $\E\approx 5.1$ TeV for $\tilde\lambda(\E)=0$.
Our results (Fig.~\ref{figyt}) are radically different from
the BHL results \footnote{Using the compositeness conditions 
$\tilde Z_{H}(\mu)=1/\bar g_t^2(\mu)\rightarrow 0, \mu\rightarrow\Lambda$, 
we reproduced the BHL results: Figures 4 and 5; Tables I and II of Ref.~\cite{bhl1990}.}, not only the completely different behaviors and numerical values of 
the Yukawa $\bar g_t(\mu)$ and quartic $\tilde \lambda(\mu)$ couplings that will be discussed in next subsection \ref{expH}, 
but also (i) the composite Higgs boson is a {\it tightly} bound state 
of $\bar t t$ pair for finite values $\tilde Z_{H}(\mu)$, 
as if it is {\it an elementary} Higgs boson; 
(ii) the phase transition from the symmetry-breaking phase to the symmetric 
phase is indicated by $\tilde \lambda(\mu)\rightarrow 0^+$ as $\mu\rightarrow \E+0^-$, will be discussed in Sec.~\ref{ptran}; 
(iii) the drastically fine-tuning (hierarchy) problem can be resolved, 
since the intermediate scale 
$\E$ ($v<\E\ll \Lambda$) replaces 
the high-energy cutoff $\Lambda$ and sets into Eq.~(\ref{gbeta}), the gap-equation
\begin{eqnarray}
\frac{1}{G_c}-\frac{1}{G}=\frac{1}{G_c}\left(\frac{m_t}{\E}\right)^2
\ln \left(\frac{\E}{m_t}\right)^2>0,
\label{delta1}
\end{eqnarray}
giving rise to a soft chiral-symmetry breaking scale $m_t\lesssim v$ 
and the pseudoscalar decay constant,
\begin{equation}
f_\pi^2
\approx \frac{N_c}{32\pi^2}m_t^2\ln \frac{\E^2}{m_t^2}=\frac{N_c}{32\pi^2}\E^2\left(1-\frac{G_c}{G}\right),
\label{decay}
\end{equation}
for more detailed discussions, see Ref.~\cite{xue2013}.

In the IR-domain ($G\gtrsim G_c$) of unique four-fermion coupling $G$, 
it seems that the 
four-fermion operators (\ref{bhlx}) undergo the SSB, leading to 
the fermion-condensation $M_{ff'} =-G\langle \bar \psi^f \psi^{f'}\rangle=m\delta_{ff'}\not=0$, 
two diagonal mass matrices of quark sectors $q=2/3$ and $q=-1/3$
satisfying $3+3$ mass-gap equations, see Eqs.~(\ref{mqu}) and (\ref{mqd}). 
It was demonstrated \cite{xue2013_1} that as an energetically favorable 
solution of the SSB ground state of the SM, only top-quark is massive 
($m^{\rm sb}_t=- G\langle \bar\psi_t \psi_t\rangle\not=0$), 
otherwise in addition to those become the longitudinal modes of massive gauge bosons, there would be more Goldstone modes contributing the 
SSB-ground-state energy.
In other words, among four-fermion operators (\ref{bhlx}) 
and (\ref{bhlxl}), the $\langle\bar t t\rangle$-condensate 
model (\ref{bhl})
is the unique channel undergoing the SSB of SM gauge symmetries, 
for the reason that this is energetically favorable, i.e., 
the ground-state energy is minimal when   
the maximal number of Goldstone modes are three and equal to the number of
the longitudinal modes of massive gauge bosons in the SM.  
Moreover,
the four-fermion operators (\ref{bhlxl}) of the lepton sector do not  
undergo the SSB leading to the lepton-condensation 
$M_{ff'} =-G\langle \bar \ell^f \ell^{f'}\rangle=m_\ell\delta_{ff'}\not=0$, i.e., 
two diagonal mass matrices of the lepton sector ($q=0$ and $q=-1$).
The reason is that the effective four-lepton coupling $(GN_c)/N_c$ is $N_c$-times smaller 
than the four-quark coupling $(GN_c)$, where the color number $N_c=3$.
In the IR-domain $(G\gtrsim G_c)$ of the IR-stable fixed point $G_c$, the effective four-quark coupling is above the critical value and 
the SSB occurs, whereas the effective four-lepton coupling is below 
the critical value and the SSB does not occur for the lepton sector. 

As a result, only the top quark acquires its mass via the SSB and
four-fermion operator (\ref{bhl}) of the top-quark channel 
becomes the relevant operator following the RG equations 
in the IR domain \cite{bhl1990}. 
While all other quarks and leptons do not acquire their masses via the SSB 
and their four-fermion operators 
(\ref{bhlx},\ref{bhlxl},\ref{bhlbv},\ref{bhlql})  
are irrelevant dimension-6 operators, 
whose tree-level amplitudes of four-fermion scatterings are highly
suppressed ${\mathcal O}[(\mu/\Lambda)^2]$, thus their deviations 
from the SM are nowadays experimentally inaccessible \cite{xue2015}. 

\subsubsection
{Experimental indications of tightly-bound composite Higgs boson ?}\label{expH}

The wave-function renormalization $\tilde Z_H(\mu)=1/\bar g_t^2(\mu)$ 
represents the form-factor of composite Higgs boson. In the BHL case, the compositeness condition $\tilde Z_H(\mu)\rightarrow 0, \mu \rightarrow \Lambda$ is adapted, this implies that the BHL loosely-bound composite Higgs boson behaves very differently from the elementary Higgs boson, and significant deviations from the SM are expected. 
In theoretical calculations, the BHL Yukawa-coupling $\bar g_t^{\rm BHL}(\mu)$ 
increases with energy $\mu$, and Higgs-quartic coupling $\tilde\lambda^{\rm BHL}(\mu)$ increases for $M_{_{Z}}<\mu \lesssim 5$ TeV (see Figs.~4 and 5 of Ref.~\cite{bhl1990}), which are completely different from our results of Fig.~\ref{figyt}. In experimental fact, the BHL results are firmly excluded by the measured values of Higgs mass and decay rate (width). 
Instead, in our case the non-vanishing 
$\tilde Z_H(\mu)$ means that after conventional wave-function and vertex 
renormalizations, the tightly-bound composite Higgs boson behaves as if an elementary particle. The deviations from the SM with an elementary Higgs boson are expected to be very small. 

However,
the effective top-Yukawa $\bar g_t(\mu)$
and Higgs-quartic $\tilde\lambda(\mu)$ couplings
monotonically and slightly decrease as the energy scale $\mu$ 
increases over the range ($M_{_Z}< \mu < \E$), see Fig.~\ref{figyt}. 
In the range $m_{_H}< \mu <\E$, 
these might imply some effects on the rates or cross-sections 
of the following three dominate processes of Higgs-boson production 
and decay \cite{ATLAS,CMS} or 
other relevant processes. Two-gluon fusion produces a Higgs boson via a top-quark loop, 
which is proportional to the effective Yukawa coupling $\bar g_t(\mu)$. Then,  
the produced Higgs boson decays into the two-photon state by coupling to a top-quark loop,
and into the four-lepton state by coupling to two massive $W$-bosons or two massive $Z$-bosons.
Due to the $\bar t\,t$-composite nature of Higgs boson, the one-particle-irreducible (1PI) 
vertexes of Higgs-boson coupling to a top-quark loop, to two massive $W$-bosons 
and to two massive $Z$-bosons are proportional to 
the effective Yukawa coupling $\bar g_t(\mu)$. 
As a result, both the Higgs-boson decaying rate $\Gamma_{H\rightarrow f}$ to each of these three channels and total decay rate 
$\Gamma^{\rm total}_H=\sum_f\Gamma_{H\rightarrow f}$ 
are proportional to $\bar g^2_t(\mu)$, however
the branching ratio ($\Gamma_{H\rightarrow f}/\Gamma^{\rm total}_H$) 
of each Higgs-decay channel is not changed. 
The energy scale $\mu$ is actually the 
Higgs-boson energy, representing the total energy of final states, 
e.g., two-photon state and four-lepton states, into which the produced Higgs boson decays. 

These discussions imply that the resonant 
amplitude (number of events) of two-photon invariant 
mass $m_{\gamma\gamma}\approx 126$ GeV and/or 
four-lepton invariant mass $m_{4l}\approx 126$ GeV is expected 
to become smaller as the produced Higgs-boson energy $\mu$ increases, i.e., the
energy of final two-photon and/or four-lepton states increases,   
when the CM energy $\sqrt{s}$ of LHC $p\,p$ collisions increases 
with a given luminosity. 
Suppose that the total decay rate or each channel 
decay rate of the SM Higgs boson is measured at the Higgs-boson energy $\mu=m_t$ 
and the SM value of Yukawa coupling 
$\bar g^2_t(m_t)=2m_t^2/v\approx 1.04$, see Fig.~\ref{figyt}. 
In the scenario of tightly-bound composite Higgs boson, 
as the Higgs-boson energy $\mu$
increases to $\mu=2m_t$, the Yukawa coupling 
$\bar g^2_t(2m_t)\approx 0.98$, 
the variation of total decay rate or each channel 
decay rate is expected to be 
$6\%$ for $\Delta \bar g^2_t\approx 0.06$. Analogously, the
variation is expected to be $9\%$ at 
$\mu=3m_t$, $\bar g^2_t(3m_t)\approx 0.95$ or $11\%$ at $\mu=4m_t$, 
$\bar g^2_t(4m_t)\approx 0.93$, as shown in Fig.~\ref{figyt}.
These variations are still too small to be clearly 
distinguished by the present LHC experiments. 
Such scenario of tightly-bound composite Higgs boson 
has not been found so far any tension with electroweak precision
variables. Nevertheless, it certainly needs more detailed analysis 
of the tightly-bound composite Higgs phenomenology in comparison with electroweak precision variables in future work.

In addition, the nonresonant new phenomena, stemming 
from four-fermion scattering amplitudes of diemnsion-6
irrelevant operators in 
effective Lagrangian (\ref{art1}), are hightly suppressed 
${\mathcal O}(\Lambda^{-2})$ and thus hard to 
be identified  \cite{xue2015} in high-energy processes of 
LHC $p\,p$ collisions (e.g., the 
Drell-Yan dilepton process, see Ref.~\cite{ATLAS2013}), 
$e^-e^+$ annihilation to hadrons and deep inelastic lepton-hadron $e^-\,p$ 
scatterings at TeV scales. Nonetheless, these effects are the nonresonant 
new signatures of low-energy collider that show the deviations 
of such scenario from the SM.

\subsubsection{Yukawa couplings and FCNC's in IR-domain}\label{irfcnc}

The SSB generated top-quark mass $m_t\propto \bar g_t v$ 
breaks chiral symmetries, 
fermion-family mixing matrices (\ref{mmud})
become relevant for coupling different fermion families. 
As a consequence, other quarks and leptons 
acquire their masses $\bar g_f v$, because Schwinger-Dyson (SD) equations 
for their self mass-energy functions acquire the explicit symmetry 
breaking (ESB) terms induced by the $W^\pm$-boson vector-like couplings 
and quark-lepton interactions at high energies, via the fermion-family 
mixing matrices like the CKM (\ref{mmud}) and PMNS matrices in the SM.  
References \cite{xue2016_1,xue2016_2} show in details how the top quark acquires its mass together with Higgs boson mass by the SSB in terms of 
$\bar g_t(\mu)$ and $\tilde\lambda(\mu)$ 
couplings (Figs.~\ref{figyt}), and other SM fermions 
acquire their masses $m_f\propto \bar g_f v$ by the ESB, 
as functions of the top-quark mass $m_t$ 
or Yukawa coupling $\bar g_t(\mu)$, 
and mixing-matrix elements of the CKM and PMNS types in the SM, 
which are intrinsic parameters fixed by experiments. The hierarchy patterns of fermion masses and mixing-matrix elements are closely related.

As a consequence, the SM chiral gauge symmetries and global 
fermion-family (flavor) symmetries are broken at a soft 
energy scale $m^f=\bar g_t v \ll \E\ll \Lambda$. The latter fermion-family 
$U_L(3)\times U_R(3)$ 
symmetries are broken to the $U(1)$ symmetry for each fermion family, i.e., $U_1(1)\times U_2(1)\times U_3(1)$. Thus the quadrilinear four-fermion operators 
(\ref{q1}), bilinear two-fermion operators (\ref{mqu}) and (\ref{mqd}) 
are ``diagonal'', namely they are for each fermion family without any
fermion-family mixing in the fermion-family space based on mass eigenstates of 
fermions. Given a definite electric charge, 
the $U(1)$-symmetry for each fermion flavor is still preserved and 
the quantum number of each fermion flavor is conserved. This completely 
prohibits any {\it four-fermion-interacting} process that violates fermion-flavor-number conservation, such as the FCNC process 
converting from one elementary fermion flavor $\psi^f$ to another $\psi^{f'}$ with the same electric charge.

However, such prohibition is relaxed as perturbatively 
turning on the interactions (\ref{art1}) 
between chiral-gauge-boson $W^\pm$ and massive fermions in addition to the four-fermion
interactions (\ref{q1}). The reason is that these chiral-gauge 
interactions are not diagonal in the fermion-family space 
based on the mass eigenstates of fermions (\ref{mqu}) and (\ref{mqd}), attributed to the fact that the mass eigenstates
of the SM {\it elementary} fermions are different from their chiral-gauge eigenstates. As a result, a 1PI-interacting vertex that
violates the fermion-flavor-symmetry $U_1(1)\times U_2(1)\times U_3(1)$ is induced via the mixing matrices of CKM 
type (\ref{mmud}) and interactions among SM gauge bosons, 
e.g.~$W^\pm$ and photon, at one-loop level.   
Such 1PI vertex allows
the FCNC processes of changing one {\it elementary} 
fermion $\psi^{f}$ flavor [mass eigenstate (\ref{cqu}) or (\ref{cqd})] to another {\it elementary} fermion $\psi^{f'}$ 
flavor [mass eigenstate] with the same electric charge. This situation 
is the same as that of the SM with an elementary Higgs boson. 

Nevertheless, such $W^\pm$-boson contributions of SM-type 
to the FCNC processes are highly suppressed, 
since they come from the loop-level contributions 
of the SM $W^\pm$-boson via the 
fermion-family mixing matrix of the CKM type (\ref{mmud}) 
in its gauge coupling vertexes.  
In conclusion, apart from the SM-type suppressed contributions to FCNC, 
the effective Lagrangian (\ref{art1}) with four-fermion operators (\ref{q1}) 
in the IR-domain does not contain any additional unsuppressed 
1PI vertexes that contribute to the FCNC processes.

\comment{
In addition, the tree-level scattering amplitudes 
of irrelevant four-fermion operators (\ref{q1}) except the 
relevant top-quark channel in the IR-domain are strongly 
suppressed [${\mathcal O}(\mu/\Lambda)^2$] by the cutoff $\Lambda$. 
}

\begin{figure}[t]
\begin{center}
\includegraphics[height=1.25in]{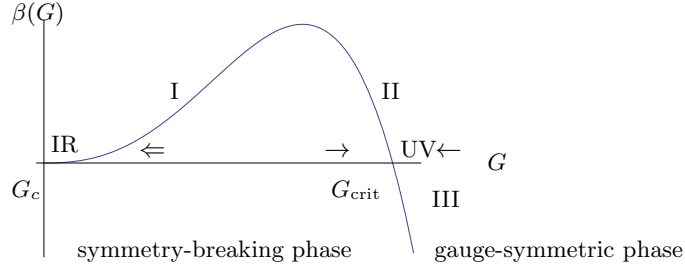}
\put(-155,90){\footnotesize $\beta(G)$}
\put(-110,38){ $\Leftarrow$ }
\put(2,38){ $\leftarrow$ }
\put(-40,38){ $\rightarrow$ }
\put(25,33){\footnotesize $G$}
\put(-130,00){\footnotesize symmetry-breaking phase}
\put(5,00){\footnotesize gauge-symmetric phase}
\put(-95,60){\footnotesize I}
\put(-15,60){\footnotesize II}
\put(4,19){\footnotesize III}
\put(-155,23){\footnotesize $G_c$}
\put(-140,40){\footnotesize ${\rm IR}$}
\put(-34,23){\footnotesize $G_{\rm crit}$}
\put(-8,38){\footnotesize ${\rm UV}$}
\caption{This is a sketch \cite{xue2014} 
to qualitatively show the behavior of the 
$\beta$-function in terms of the four-fermion coupling $G$. We indicate 
the IR-stable fixed point $G_c$ and a possible UV-stable fixed 
point $G_{\rm crit}$, the latter separates the symmetry-breaking phase 
(positive $\beta(G)$-function) from the gauge-symmetric phase (negative 
$\beta(G)$-function). We also indicate the different parts of 
$\beta(G)$-function: the positively increasing part ``I'' and the IR-domain where RG flow  approaches (thick left arrow) to the IR-fixed point, as energy scale $\mu$ decreases; (ii) the 
positively decreasing part ``II'' and the negative part ``III'', 
as well as the UV-domain where RG flow  approaches (thin right 
and left arrows) to the UV-fixed point, as the running energy scale $\mu$ increases.} \label{fixp}
\end{center}
\end{figure} 

\subsection{Gauge-symmetric phase of strong four-fermion coupling}\label{uvds}

We turn to the  brief recall on
the gauge-symmetric phase of strong four-fermion coupling $G$,
\begin{equation}
-G\bar\psi_{_L}\psi_{_R}\bar\psi_{_R} \psi_{_L},
\label{4foa}
\end{equation}
preserving $SU_L(2)$ gauge symmetry, where two-component Weyl fermion $\psi_{_L}$ ($\psi_{_R}$) is an $SU_L(2)$ doublet (singlet).
The strong-coupling expansion was adopted to calculate 
two-point functions of composite boson and fermion fields \cite{xue1997}, and the vertex functions of their couplings to the $SU_L(2)$-gauge bosons \cite{xue2000,xue2000prd,xue2003}. Detailed calculations can be found 
in these references and their appendixes. 
In the lowest non-trivial order, 
we obtained the propagators of
the massive composite bosons ${\mathcal A}=(\bar \psi_{R} \psi_L)$ and composite Dirac fermions: ${\Psi}_D=(\psi_L,{\mathcal A}\psi_R)$ and 
${\Psi}_S=({\mathcal A}^\dagger\psi_L,\psi_R)$,
\begin{equation}
\langle{\mathcal A}(0) {\mathcal A}^\dagger(x)\rangle \Rightarrow
\frac{Z^{^S}_B}{q^2+M^2_B};\quad \langle\Psi_{D,S}(0) \bar\Psi_{D,S}(x)\rangle
\Rightarrow Z^{^S}_{D,S}
\frac{ip^\mu\gamma_\mu +M_{D,S}}{p^2+M^2_{D,S}},
\label{4fo}
\end{equation}
where the pole $M_{B,D,S}$ and residue $Z^{^S}_{B,D,S}$ respectively
represent mass and form factor of composite particles. As long as their form-factors are finite, these composite particles behave as elementary particles. The propagators of renormalized composite particles (\ref{4fo}) give their mass-shell conditions,
\begin{eqnarray}
E_B=\sqrt{q^2+M^2_B}\approx M_B,\quad E_{D,S}=\sqrt{p^2+M^2_{D,S}}\approx M_{D,S},\quad {\rm for }~~q^2,p^2\ll M^2_{B,D,S}.
\label{mcp}
\end{eqnarray}
The vertex functions of their $SU_L(2)$-couplings were obtained by Ward identities from propagators (\ref{4fo}), 
because the composite-particle spectra preserve 
the $SU_L(2)$-gauge symmetry.   

These massive composite particles formed by the such 
strong-coupling dynamics in the gauge-symmetric phase are completely different from the composite particles like massive Higgs boson and massless Goldstone bosons formed by the weak-coupling NJL dynamics 
in the symmetry-breaking phase where the SSB takes place. 

\subsection{Strong critical coupling and the UV-stable fixed point}\label{uvd}
This indicates that there are two distinct phases: 
(i) the symmetry-breaking phase 
$G_c<G<G_{\rm crit}$ for the SM of elementray particles; (ii) the 
gauge-symmetric phase $G> G_{\rm crit}$ for an effective theory of composite particles, and the second-order phase transition from one phase to another at the strong critical coupling $G_{\rm crit}$. 

\subsubsection{Strong critical coupling}\label{ptran}

In the symmetric phase, we indeed found \cite{xue1997} the existence of 
strong critical coupling $G_{\rm crit}$ by using strong four-fermion coupling expansion to approximately calculate the two-point Green function (\ref{4fo}) of composite boson and examine sign change of quadratic term $M^2_B{\mathcal A}{\mathcal A}^\dagger(x)$.
Instead, based on the symmetry-breaking phase of weak four-fermion coupling, see Sec.~\ref{ird}, we show 
the existence of strong critical point $G_{\rm crit}$ 
of the second-order phase transition from the symmetry-breaking phase to 
gauge-symmetric phases. 

In Refs.~\cite{xue2013,xue2014}, 
we solve the full one-loop RG 
equations \cite{bhl1990} for running couplings $\bar g_t(\mu^2)$ and $\bar \lambda(\mu^2)$
with the top-quark and Higgs-boson mass-shell conditions
\begin{eqnarray}
m_t=\bar g_t(m_t)v/\sqrt{2},\quad m_{_H}^2/2= \tilde \lambda (m_{_H}) v^2,
\label{thmass}
\end{eqnarray}
$ m_t\approx 172.9\, {\rm GeV}$ and
$m_{_H}\approx 126 \,{\rm GeV}$ fixed by experiments.
As a result, we uniquely determine 
the renormalized form factor $\tilde Z_{H}(\mu)$ 
and quartic coupling $\bar\lambda(\mu)$ of the composite Higgs particle. We also find the intermediate energy scale $\E$, at which the effective quartic coupling $\tilde\lambda(\E)=0$, and the finite value of form-factor
$\tilde Z_H(\E)$,
\begin{eqnarray}
\E \approx  5.1\,\, {\rm TeV},\quad \tilde Z_H(\E)\approx 1.26,\quad \tilde\lambda(\E)=0,
\label{thvari}
\end{eqnarray}
otherwise the effective
theory would run into an instability ($\tilde \lambda\sim 0^-$) 
beyond $\E$. 
These are certainly preliminary and qualitative results, since we use 
one-loop RG equations in low-energy and weak-coupling region and extrapolate their solutions to high-energy and strong-coupling region.   
Nevertheless, the solution (\ref{thvari}) implies the following three important features. (i) In the effective Lagrangian of quadratic 
term $m^2_{_H}HH^\dagger/2$, 
the squared Higgs-boson mass $m^2_{_H}=2\tilde\lambda(\mu) v^2$
changes its sign at $\mu=\E$, indicating the second-order 
phase transition $G_{\rm crit}$ from the symmetry-breaking phase to a gauge-symmetric phase of high-energy and strong-coupling region. 
(ii) The form-factor $\tilde Z_H(\mu)\not=0$ shows that 
the tightly bound composite Higgs particle behaves 
as if an elementary particle for $\mu\leq\E$. 
(iii) The effective form-factor $\tilde Z_H(\E)$ 
of composite Higgs boson is finite, implying the formation of massive composite fermions 
$\Psi\sim (H\psi$) in the gauge-symmetric phase. 

To close this subsection, it is worthwhile to mention 
that in Ref.~\cite{DCW2015} 
it is shown in the elementary Higgs-boson model that the quadratic 
term from high-order quantum corrections has a physical 
impact on the SSB and the phase transition to a symmetric phase occurs 
at the scale of order of TeV.

\subsubsection{UV-stable fixed point}

In order to show that this critical point $G_{\rm crit}$ of the second-order phase transition can be a UV fixed point,   
we calculated the $\beta$-function in the symmetric phase of strong 
four-fermion coupling. Up to
the lowest non-trivial order of the strong coupling expansion, 
i.e., the two-fermion-loop contribution (sun-set diagrams),
and we obtained the negative $\beta$-function \cite{xue2014}
\begin{eqnarray}
\beta(G) = p^2\frac{\partial {\mathcal G}(p)}{\partial p^2}\approx \frac{6}{{\mathcal G}} \frac{\partial \Phi(p^2/\Lambda^2)}{\partial \ln(p^2/\Lambda^2)}< 0 ,
\label{ren}
\end{eqnarray}
where ${\mathcal G}\equiv G\times (\Lambda/\pi)^2$
and the dimensionless Lorentz-scalar function $\Phi(p^2/\Lambda^2)$ is 
positive and finite, monotonically decreases as 
the energy scale $p^2/\Lambda^2$ increases.
On the basis of the $\beta(G)$-function being positive (\ref{gbeta}) 
and becoming negative (\ref{ren}), as sketched as ``I'', ``II'' 
and ``III'' in 
Fig.~\ref{fixp}, we infer there must be at least one 
zero-point of the $\beta(G)$-function, i.e., $\beta(G_{\rm zero})=0$ and 
$\beta'(G_{\rm zero})<0$. This zero-point $G_{\rm zero}$ is a UV-stable 
fixed point, which should coincide with the strong critical
coupling $G_{\rm zero}\approx G_{\rm crit}$ for the reason that a UV-stable fixed point should be the candidate of the critical point 
$G_{\rm crit}$ for the second-order phase transition from the 
symmetry-breaking phase to the symmetric phase. 
As the running energy scale $\mu$ increases, the effective coupling 
$G(\mu)\rightarrow G_{\rm crit}\pm 0$ (thin right 
and left arrows, see Fig.~\ref{fixp}) is attracted to this UV-stable fixed point from both symmetry-breaking and symmetric phases.
In the scaling region (vicinity) of UV-stable fixed point $G_{\rm crit}$, 
i.e., UV-domain, RG flow approaches to the UV-fixed point,  as energy scale $\mu$ increases, in the scenes of renormalization group 
(scaling) invariance and resultant RG equations.

This is analogous to non-linear $\sigma$ models \cite{GML}, which contain a dimensionful coupling constant and are thus not perturbatively renormalizable. Nevertheless, the nonperturbative critical coupling of 
these $\sigma$ models shows the second-order phase transition from the 
ordered (symmetry-breaking) phase to the disordered (symmetric) phase, 
and exhibits a non-trivial UV-fixed point and UV-scaling domain of the renormalization group both in the lattice formulation \cite{sigmaL} and in the $2+\epsilon$ formulation \cite{sigmaE,Kleinertsigma}. 

However, due to the lack of a non-perturbative method to effectively 
approach to the critical coupling $G_{\rm crit}$ and its neighborhood, 
we have not been able to quantitatively determine 
the $G_{\rm crit}$ value
and properties of its scaling region 
i.e., the UV-domain, which will be qualitatively discussed and recalled
in next section.

\subsection{The UV-domain of UV-stable fixed point}\label{uvd1}

\subsubsection{Energy threshold of composite particles}

As the running energy scale $\mu$ decreases in the vicinity of UV-stable fixed point $G_{\rm crit}$, the $\beta(\mu)$ function 
(see Fig.~\ref{fixp}) shows that the RG flows take the effective theory of composite particles away from the UV fixed point towards 
the IR-domain of the IR-fixed point $G_c$, where the low-energy SM of elementary particle physics is realized. This implies the existence of 
the energy threshold $\E_{\rm thre}$, below which $\mu < \E_{\rm thre}$
composite particle dissolves into its constituents of SM elementary particles. As discussed in Sections V and VI of Ref.~\cite{xue2000prd}, when the energy scale $\mu$ decreases to the energy threshold 
$\E_{\rm thre}$ and $G(\mu)\rightarrow G_{\rm crit}(\E_{\rm thre})$, 
the phase transition occurs from the symmetric phase to the 
symmetry-breaking phase, all composite particles (poles) dissolve into their constituents, which are represented by three-fermion, fermion-boson 
and two-fermion cuts in the energy-momentum plane, as their form factor and binding energy vanish \cite{weinberg}. Actually, the energy threshold
$\E_{\rm thre}$ represents the symmetry-breaking scale at the second-order phase transition $G_{\rm crit}$. 

The composite-particle masses 
$M$ (\ref{mcp}) contain the negative binding energy 
$-{\mathcal B}[G(\mu)]$
and positive kinetic energies ${\mathcal K}$ of their constituents. The energy threshold 
$\E_{\rm thre}$ is determined by 
${\mathcal B}[G(\mu)]_{\mu\rightarrow\E_{\rm thre}}\rightarrow {\mathcal K}$ and vanishing form factors of composite particles in the 
symmetric phase. These calculations require non-perturbative approaches that need to be developed in future.  
Nevertheless, we gain some physical insight into 
the symmetry-breaking scale $\E_{\rm thre}\gtrsim \E \approx 5.1$ GeV, since the approximate $\E$-value (\ref{thvari}) is obtained in symmetry-breaking phase by extrapolating the solutions of the one-loop 
RG-equations of low-energy ($v$) region to high-energy ($\E$) region.  

\subsubsection{Characterisic energy scale}

In the IR-domain of the fixed point $G_c$ where the effective Lagrangian 
of the SM is realized, the correlation length $\xi_v$ corresponding to
the electroweak scale $v=239.5\,$GeV 
that is attributed to the SSB dynamics 
in symmetry-breaking phase sets the physical energy 
scale of IR-domain. This characteristic energy scale is much smaller than the cutoff ($v\ll \Lambda$), leading to the renormalization group (scaling) invariance and resultant RG equations in the IR-domain.   
This characteristic electroweak scale $v$  
is determined by the experimental measurements 
of intermediate gauge bosons $W^\pm$ and $Z^0$ masses. 

Analogously, as discussed in Ref.~\cite{xue2014}, in the UV-domain (scaling region) of the UV-fixed point (critical point) $G_{\rm crit}$ of the second-order phase transition, 
the correlation length $\xi$ or characteristic energy  
$\E_\xi=\xi^{-1}$ sets the physical scale. 
The correlation length $\xi$ is much larger than the cutoff 
$a\approx \Lambda^{-1}$, i.e., 
$\xi\gg a$ or $\E_\xi\ll \Lambda$ leads to the renormalization group (scaling) invariance and resultant RG equations in the UV-domain. In this scaling domain, the running coupling 
$G(a/\xi)$ can be expanded as a series,
\begin{eqnarray}
G(a/\xi)&=& G_{\rm crit} \left[1+ a_0(a/\xi)^{1/\nu}+ {\mathcal O}[(a/\xi)^{2/\nu}]\right]\rightarrow G_{\rm crit}+0^+,
\label{gexp}
\end{eqnarray}
for $a/\xi\ll 1$, leading to the $\beta$-function 
\begin{eqnarray}
\beta(G)&=& (-1/\nu) (G-G_{\rm crit}) + {\mathcal O}[(G-G_{\rm crit})^2]<0\,.
\label{betaexp}
\end{eqnarray}
The correlation length $\xi$ follows the scaling law
\begin{eqnarray}
\xi &= & c_0 a\exp \int^G\frac{dG'}{\beta(G')}
=\frac{ c_0a}{(G-G_{\rm crit})^\nu},
\label{xivary}
\end{eqnarray}
where the coefficient $c_0=(a_0G_{\rm crit})^\nu$ and critical 
exponent $\nu$ need to be determined by 
non-perturbative numerical simulations. 
The physical scale $\E_\xi\equiv \xi^{-1}$ in the UV-domain 
is attributed to the strong-coupling dynamics of forming composite particles. This implies the masses of composite particles (\ref{4fo}) 
and (\ref{mcp})
\begin{eqnarray}
M \propto \E_\xi=\xi^{-1},
\label{mcom}
\end{eqnarray} 
and the running coupling  
$G(\mu)|_{\mu\rightarrow\E_{\rm thre}+0^+}\rightarrow G_{\rm crit}$, 
\begin{eqnarray}
G(\mu)\simeq G_{\rm crit}\left[1-\frac{1}{\nu}\ln\left(\frac{\mu}{\E_\xi}\right)\right]^{-1}, \quad \mu/\E_\xi=\xi/(aa_0^\nu) >1\,,
\label{runnc}
\end{eqnarray}
and the scale $\mu$ indicates the energy transfer between constituents 
inside composite particles.

On the basis of these discussions and observations, we advocate the 
following relation for 
(i) the energy scale $\E\approx 5\,$ TeV of Eq.~(\ref{thvari}) 
extrapolated by the RG equations in the IR-domain for the SM, 
(ii) the energy threshold $\E_{\rm thre}$ corresponding to the phase-transition scale from symmetric to symmetry-breaking phases and (iii) the characteristic 
energy scale $\E_\xi$ in the UV-domain for the effective theory of composite particles
\begin{eqnarray}
\E\lesssim \E_{\rm thre} \lesssim \E_\xi\ll \Lambda, \quad \E\approx 5\, {\rm TeV}.
\label{scales}
\end{eqnarray}
The values of these characteristic scales of 
UV-domain need some experimental knowledge in high energies, 
analogously to the electroweak scale ($v$) of IR-domain. 

\subsection{ Relevant and irrelevant operators in IR- and UV-domains}\label{riro}
\hskip0.1cm 
In the weak-coupling IR-domain, as an energetically favorable solution 
for the SSB ground state \cite{xue2013_1}, among four-fermion operators 
(\ref{art1}) the top-quark channel 
$G(\bar\psi^{ia}_Lt_{Ra})(\bar t^b_{R}\psi_{Lib})$ \cite{bhl1990} 
is the only physically relevant and renormalizable operators of effective dimension-4, due to the NJL-dynamics for the SSB. Namely, it becomes 
the effective SM Lagrangian (\ref{bhl}) 
with {\it bilinear} top-quark mass term and 
Yukawa coupling to the composite Higgs boson $H$, 
which obeys the RG equation approaching to
the low-energy SM physics characterized by the energy 
scale $v\approx 239.5$ GeV. 
Other four-fermion operators in Eq.~(\ref{art1}), as well as repulsive four-fermion operators (\ref{art0}), 
do not undergo the SSB and are irrelevant
dimension-6 operators, whose tree-level amplitudes of four-fermion scatterings are suppressed ${\mathcal O}(\Lambda^{-2})$, thus their deviations from the SM could not be experimentally accessible today.

In the strong-coupling UV-domain,
all attractive four-fermion operators (\ref{art1}) are physically relevant operators associating to the strong-coupling dynamics 
and formation of composite bosons and fermions (\ref{4fo}). 
Due to the unique four-fermion coupling $G$, 
this strong-coupling dynamics of forming composite particles occurs 
in all channels of Eq.~(\ref{art1}) for each SM fermion family.
These composite particles on mass-shells behave as if they were elementary, as long as their form factors are finite.
All 1PI functions $\Gamma[\mu, G(\mu)]$ 
of the quantum field theory 
(\ref{art1}) at the UV-cutoff scale $\Lambda$ 
evolve to either irrelevant or relevant 1PI functions,
as the running energy scale $\mu$ increases. The irrelevant 1PI 
functions are suppressed by powers of $(\E_\xi/\Lambda)^n$ and thus 
decouple from the effective theory. Instead, the relevant 1PI
functions follow the scaling law (RG equations), therefore are 
effectively dimension-4 and renormalizable, for example 
the propagators of composite fermions and bosons and 
their vector-like coupling vertexes to the SM gauge 
bosons.   
The effective field theory in this UV-domain is expected to 
contains the massive composite 
particle spectrum preserving the SM chiral gauge symmetries, and
relevant operators of effective dimension-4 following the RG equations.
Compared with the SM in the IR-domain, the effective 
theory of composite particles in 
the UV-domain has the same chiral gauge symmetries (quantum 
numbers) and couplings to gauge bosons ($\gamma,W^\pm, Z^0$ and gluon), but the different vector-like spectra and coupling vertexes, 
apart from massive composite particles being comprised by SM elementary particles. Some of these peculiar features and their   
possible experimental implications were studied in 
Refs.~\cite{xue2014,xue2015,xue2016_1}. 
These are the properties of effective composite-particle 
theory in the UV-domain that we attempt to study in this article 
and in future.
 
To end this section, it is worthwhile to note that the repulsive four-fermion 
operators (\ref{art0}) are irrelevant operators of dimension-6, and 
thus suppressed ${\mathcal O}(\Lambda^{-2})$ for the reasons that 
they are neither associated with the NJL dynamics 
of the SSB in the IR-domain, nor associated 
with the strong-coupling dynamics of forming composite particles 
in the UV-domain. 

It is also worthwhile to mention that 
these discussions are reminiscent of the 
asymptotic safety \cite{w1} that quantum 
field theories regularized at UV-cutoff $\Lambda$ 
might have a non-trivial 
UV-stable fixed point, RG flows 
are attracted into the UV-stable fixed point  
with a finite number of physically renormalizable operators. 
The weak and strong four-fermion coupling $G$ brings us into 
two distinct domains. 
This lets us also recall the quantum chromodynamics (QCD): 
asymptotically free quark states 
in the domain of a UV-stable fixed point
and bound hadron states in the domain of 
a possible IR-stable fixed point \footnote{
The references on this issue can be found in the recent paper:
Z.-Y. Zheng and G.-L.  Zhou, ``A method for getting well-defined coupling constant in all region '',  
arXiv:1705.07430
}
.

\section
{\bf Composite particles and effective Lagrangian in UV-domain}
\label{composite}

In the gauge symmetric phase of an $SU_L(2)$-chiral gauge theory containing the four-fermion operator (\ref{4foa}), 
the composite particles and 1PI vertex functions, 
as well as their properties in the UV-domain were 
analyzed by using the approach of 
strong-coupling expansion to the dynamics of four-fermion operators 
at the cutoff $\Lambda$ \cite{xue1997,xue2000prd, xue2003, xue2014,xue2015}. In this section, however, based on the effective Einstein-Cartan Lagrangian (\ref{art1}) in the SM framework, see Sec.~\ref{sm4f}, 
we are going to show 
the composite-particle spectrum and 1PI interaction 
(effective Lagrangian) in terms of the quantum numbers of SM gauge symmetries. The purpose is that we shall go further to discuss composite particles 
decay and other processes into SM elementary gauge bosons and fermions, relevantly to possible high-energy experiments.
These are the main results of this article presented in this section and 
following sections \ref{cbd}, \ref{cfc} and \ref{spec}. 

\subsection
{Composite particles in UV-domain}\label{composites}

\subsubsection{Quark sector}
  
Performing strong-coupling calculations similar to those detailedly
presented in the Ref.~\cite{xue1997} and their appendixes, 
we obtain the following results. For the $u$-quark channel, 
the massive composite boson is an $SU_L(2)$-doublet 
\begin{equation}
{\mathcal A}^i=[Z^{^S}_\Pi]^{-1/2}(\bar u_{Ra} \psi^{ia}_L)
\label{boundb}
\end{equation}
of the hypercharge $Y=-1/2$ and $i (a)$ being the weak isospin (color) 
index. This composite boson
combines with an elementary two-component Weyl quark to form the composite two-component Weyl-fermion states (three-fermion states),
\begin{equation}
{\bf\Psi}^{ib}_R=(Z^{^S}_R)^{-1}{\mathcal A}^iu^b_{R}\,;
\quad {\bf\Psi}^b_{L}=(Z^{^S}_L)^{-1}{\mathcal A}^{i\dagger}\psi^{b}_{iL},
\label{bound}
\end{equation}
and to form the massive composite four-component Dirac-fermion states: 
the $SU_L(2)$ doublet ${\bf\Psi}^{ib}_D$ and 
the $SU_L(2)$ singlet ${\bf\Psi}^{b}_S$ \footnote{ 
A four-component composite 
Dirac fermion composes an elementary Weyl fermion 
and a composite Weyl fermion, each of them is two-component, either 
left- or right-handed (a definite chirality ) \cite{xue1997},  
similar discussions for massive composite bosons,  
see  Ref.~\cite{xue2014}.}, 
\begin{equation}
{\bf\Psi}^{ib}_D=(\psi^{ib}_L, {\bf\Psi}^{ib}_R),\qquad 
{\bf\Psi}^{b}_S=({\bf\Psi}^b_{L},u_R^b).
\label{boundd}
\end{equation}
The form factors $[Z^{^S}_\Pi]^{1/2}$ (\ref{boundb}) and 
$Z^{^S}_{R,L}$ (\ref{bound}) are 
generalized wave-function renormalizations 
of the composite boson and
fermion operators.

The four-component composite Dirac fermions are vector-like spectra, 
fully preserving the parity symmetry. 
The composite $SU_L(2)$ doublets ${\bf\Psi}^{ib}_D$ carry the same $U_Y(1)$-hypercharge 
as the SM elementary and left-handed $SU_L(2)$ doublets ${\psi}^{ib}_L$. 
The composite $SU_L(2)$ singlets ${\bf\Psi}^{b}_S$ carry the same $U_Y(1)$-hypercharge 
as the SM elementary and right-handed $SU_L(2)$ singlets $u^{b}_R$. 
These composite particles have the SM-gauge symmetric masses
\begin{eqnarray}
M^2_\Pi{\mathcal A}^{i\dagger}{\mathcal A}^{i},\quad
M_F\bar{\bf\Psi}^{ib}_D{\bf\Psi}^{ib}_D,\quad {\rm and}\quad 
M_F\bar{\bf\Psi}^{b}_S{\bf\Psi}^{b}_S.
\label{boundm}
\end{eqnarray}  
For the $d$-quark channel, the composite particles are represented by 
Eqs.~(\ref{boundb}-\ref{boundd}) 
with the replacement $u_{Ra}\rightarrow d_{Ra}$. In this case,
${\mathcal A}^i$ has the $U_Y(1)$-hypercharge $Y=1/2$. ${\bf\Psi}^{b}_L$ and 
${\bf\Psi}^{b}_S$ have the same $U_Y(1)$-hypercharge as the quark field 
$d^{b}_R$. 
 ${\bf\Psi}^{ib}_R$ and ${\bf\Psi}^{ib}_D$ 
have the same $U_Y(1)$-hypercharge, i.e., the $U_Y(1)$-hypercharge of the SM quark fields ${\psi}^{ib}_L$.

This represents the first family of the 
composite particles composed by 
the SM elementary particles (quarks) in the first family. 
The composite boson and fermion states (\ref{boundb}-\ref{boundd})
are the eigenstates of both the SM gauge interactions and mass 
operators (\ref{boundm}), due to the SM gauge symmetries are preserved by the massive spectra of composite particles. 
The same discussions also apply for the second and third quark families 
by substituting the $SU_L(2)$ doublet $(u_{La}, d_{La})$ 
into $(c_{La}, s_{La})$ or $(t_{La}, b_{La})$  and singlet 
$u_{Ra}$ into $t_{Ra}$ or $c_{Ra}$, as well as singlet 
$d_{Ra}$ into $b_{Ra}$ or $s_{Ra}$. 

\subsubsection{Lepton sector}

In the lepton sector, the composite boson and Weyl-fermion states formed by the first 
term ($\ell_R$-channel) of Eq.~(\ref{bhlxl}) are:
\begin{eqnarray}
{\mathcal A}^i=[Z^{^S}_H]^{-1/2}(\bar\ell_{R} \ell^{i}_L);
\qquad {\bf\Psi}^{i}_R=[Z^{^S}_R]^{-1}{\mathcal A}^i\ell_{R}\,,
\quad {\bf\Psi}_{L}=[Z^{^S}_L]^{-1}{\mathcal A}^{i\dagger}\ell_{iL},
\label{boundl}
\end{eqnarray}
and the massive composite Dirac fermions ${\bf\Psi}^{i}_D=(\ell^{i}_L, {\bf\Psi}^{i}_R)$ 
and ${\bf\Psi}_S=({\bf\Psi}_{L},\ell_R)$. The composite boson doublets 
${\mathcal A}^i=[Z^{^S}_H]^{1/2}(\bar\ell_{R} \ell^{i}_L)$ carry 
the hypercharge $Y=1/2$, the doublets 
${\bf\Psi}^{i}_D=(\ell^{i}_L, {\bf\Psi}^{i}_R)$ carry the $U_Y(1)$ hypercharge of 
the SM field $\ell^{i}_L$ 
and the singlets ${\bf\Psi}_S=({\bf\Psi}_{L},\ell_R)$ carry 
the $U_Y(1)$ hypercharge of the SM field $\ell_R$. Compared with their counterparts 
(\ref{boundb}-\ref{boundd}) in the quark sector, 
these massive composite particles have different 
quantum numbers of the SM gauge symmetries. 
The composite particles formed by the second 
term ($\nu^\ell_R$-channel) of Eq.~(\ref{bhlxl}) are obtained by Eq.~(\ref{boundl})
with the replacement $\ell_R\rightarrow \nu^\ell_R$. In this case,
${\mathcal A}^i$ has the $U_Y(1)$-hypercharge $Y=-1/2$. ${\bf\Psi}_L$ and ${\bf\Psi}_S$ have 
the $U_Y(1)$ hypercharge of the right-handed neutrino $\nu_R$.
${\bf\Psi}^{i}_L$ and ${\bf\Psi}^{i}_D$ have the same $U_Y(1)$-hypercharge, i.e., the $U_Y(1)$ hypercharge of the 
SM lepton field  $\ell^{i}_L$.

\subsubsection{Quark-lepton sector}\label{iql}

Analogously, we present for the
$d^{\,a}_{R}$- and $e_R$-channel of quark-lepton 
interactions (\ref{bhlql}), 
the massive composite Dirac fermions: $SU_L(2)
$ doublet ${\bf\Psi}^{i}_D=(\ell^{i}_L, {\bf\Psi}^{i}_R)$ and singlet 
${\bf\Psi}_S=({\bf\Psi}_{L},e_R)$, where the renormalized composite boson and composite Weyl-fermion states are:
\begin{equation}
{\mathcal A}^i=[Z^{^S}_H]^{-1/2}(\bar d^{\,a}_{R}\psi^{i}_{La});\quad
{\bf\Psi}^{i}_R=[Z^{^S}_R]^{-1}{\mathcal A}^i e_{R}\,,
\quad {\bf\Psi}_{L}=[Z^{^S}_L]^{-1}{\mathcal A}^{i\dagger}\ell_{iL},
\label{boundem}
\end{equation}
which respectively carry the $U_Y(1)$-hypercharge: $Y=1/2$, $Y=-1/2$ and $Y=-1$. 
For the $\nu^{\,e}_R$-channel, the composite particles are represented 
by Eq.~(\ref{boundem}) with the replacements 
$e_{R}\rightarrow \nu^{\,e}_{R}$ and $d^{\,a}_{R}\rightarrow u^{\,a}_{R}$. 
In this case, the counterparts of the 
composite states (\ref{boundem})
respectively carry the $U_Y(1)$-hypercharge: $Y=-1/2$, $Y=-1/2$ and $Y=0$. 
\comment{
\begin{equation}
{\mathcal A}^i=[Z^{^S}_H]^{-1/2}(\bar u^{\,a}_{R}\psi^{i}_{La});\quad
{\bf\Psi}^{i}_R=(Z^{^S}_R)^{-1}{\mathcal A}^i \nu^{\,e}_{R}\,;
\quad {\bf\Psi}_{L}=(Z^{^S}_L)^{-1}{\mathcal A}^{i\dagger}_R\ell_{iL},
\label{boundenu}
\end{equation}
}
 
The composite particles from the second and third lepton families 
can be obtained by substitutions: 
$e\rightarrow \mu, \tau$, $\nu^e\rightarrow \nu^\mu, \nu^\tau$, and 
$u\rightarrow c, t$ and $d\rightarrow s, b$. In general, 
the $U_Y(1)$-hypercharge $Y$ and $U_{\rm em}$-electric charge $Q_i$
of the composite particle
is the sum of its constituents' hypercharges
and electric charges, obeying the relation $Q_i=Y+t_{3L}^i$ in units of $e$, where $t_{3L}^i$ is 
the diagonal third component of $SU_L(2)$-isospin, 
$t_{3L}^1=1/2$ for the neutrino and up quarks, and $t_{3L}^2=-1/2$ for the electron 
and down quarks. 

\subsubsection{Discussions and three-family replication}

In the quark and lepton sectors, the massive composite boson ${\mathcal A}^i$, 
the massive composite Dirac fermions $SU_L(2)$ doublet $\Psi^i_D$ 
and singlet $\Psi_S$ carry 
an electric charge $Q=(2/3,-1/3,-1,0)$ for $u_R^a$- and $d_R^a$-quark channels, $e_R$- and $\nu_R$-lepton channels respectively. 
In the quark-lepton sector,
the massive composite boson ${\mathcal A}^i$, 
the massive composite Dirac fermions $SU_L(2)$ doublet $\Psi^i_D$ 
and singlet $\Psi_S$ carry 
an electric charge $Q=(2/3,-1/3,-1,0)$ for the $d_R^a$-$e_R$ channel
and $u_R^a$-$\nu_R$ channel respectively. It should be mentioned that
the composite bosons ${\mathcal A}^i$ carry neither the baryon number nor the lepton 
number, whereas the composite fermions $\Psi_D^i$ and $\Psi_S$ carry the baryon 
or the lepton number of the SM elementary fermions.

These first-family composite particles are composed only by the SM elementary particles in the first family, and 
there are no extra elementary gauge bosons and fermions, 
except the right-handed sterile neutrino $\nu_R$. 
The second-family (third-family) composite particles are composed only by the SM elementary particles in the second (third) family. The three families 
of composite particles are replicated corresponding to the three 
families of elementary fermions in the SM.

Though composite particles are massive, they carry the quantum numbers 
of the SM chiral gauge symmetries, which are the sum of quantum numbers 
of their constituents of the SM  elementary particles. 
The propagators of these composite particles have 
poles and residues that respectively represent their masses and 
form factors \cite{xue1997,xue2000,xue2000prd}.  
As long as their form factors are finite, 
these composite particles behave as elementary particles. It should be mentioned that 
the gauge symmetric masses $M_{\Pi,F}$ and form factors $Z^S_{\Pi,L,R}$ 
of composite particles, e.g., (\ref{boundb}-\ref{boundm}) can be different 
from one to another, due to some other effects that we do not study here. 

To end this section, we present some discussions 
on the third-family composite 
bosons (\ref{boundb}) by the top and bottom quarks, 
compared with the composite Higgs and 
Goldstone bosons. The former comprises of the massive bound states 
$(\bar t t)^s,(\bar t\gamma_5t)^s$ and $(\bar t\gamma_5b)^s$ 
formed by the strong-coupling dynamics 
in the UV-domain of the symmetric phase. They are different from  
the composite massive Higgs boson $H=(\bar t t)$ and massless 
Goldstone bosons $(\bar t\gamma_5t)$ 
and $(\bar t\gamma_5b)$ \cite{bhl1990} 
formed by the NJL dynamics in the IR-domain of the 
symmetry-breaking phase. Nevertheless, 
at the strong critical point $G_{\rm crit}$ of second-order 
phase transition that may occur around the scale $\E_\xi$, 
see Eq.~(\ref{thvari}), from the point view of the different 
ground states of the symmetric phase and
the symmetry-breaking phase, we expect some 
relationships of the binding energies and form factors 
between two types of composite bosons, e.g.,  
$(\bar t t)^s$ and $(\bar t t)$, with the same SM quantum numbers.
We will study this issue and its experimental implications in future.    

\subsection{Effective Lagrangian of composite particles in UV-domain}\label{efflag}

\subsubsection{SM effective Lagrangian of elementary particles in IR-domain}

To compare and contrast with the effective Lagrangian of composite particles 
in the UV-domain of the symmetric phase, we also present the effective SM Lagrangian  
in the IR-domain of the symmetry-breaking phase. The well-studied 
SM effective Lagrangian of elementary fermions in the first family reads
\begin{eqnarray}
{\mathcal L}_{\rm SM}&=& \sum_i \bar\psi^{i}(i\gamma_\mu\partial_\mu - m_i-m_iH/v)\psi^{i}-
e\sum_iQ_i\bar\psi^{i}\gamma^\mu\psi^{i}A_\mu\nonumber\\
&-& \frac{g_2}{2\sqrt{2}}\sum_i\bar\psi^{i}\gamma^\mu(1-\gamma_5)(T^+W^+_\mu+T^-W^-_\mu)
\psi^{i} \nonumber\\
&-&\frac{g_2}{2\cos\theta_W}\sum_i\bar\psi^{i}\gamma^\mu(g_V^i-g_A^i\gamma_5)
\psi^{i}Z_\mu ~+~{\rm Higgs ~~sector},
\label{smf}
\end{eqnarray}
for the both quark and lepton sector in one SM family, where
$i$ is the $SU_L(2)$-isospin index, 
$T^\pm$ is the weak isospin raising and lowering operators
$g_V^i=t^i_{3L}-Q_i\sin^2 \theta_W$ and $g_A^i=t^i_{3L}$.
The weak angle $\theta_W=\tan^{-1}(g_1/g_2)$ and the electron 
electric charge $e=g_2\sin\theta_W$.
These massive fermions couple to the massless photon $A_\mu$, 
massive intermediate gauge bosons $W^\pm_\mu$ and $Z_\mu$. 
The gauge and mass eigenstates of the SM elementary fermions are different, 
leading to the fermion-family mixing, when three fermion families are considered. 
The phenomenological Yukawa couplings $\bar g_i=m_i/v$ of Higgs boson and elementray 
fermion $\psi_i$ of mass $m_i$ are related to the top-quark Yukawa coupling 
$\bar g_t$ of Eq.~(\ref{bhl}) discussed in Sec.~\ref{ird}, and see Refs.~\cite{xue2016_1,xue2016_2} 
for some details.

\subsubsection{Effective Lagrangian of composite particles in UV-domain}

In the UV-domain of the symmetric phase, massive composite bosons and fermions (see Sec.~\ref{composites}) have the quantum numbers of SM chiral gauge symmetries 
$SU_c(3)\times SU_L(2)\times U_Y(1)$, couplings $g_1,g_2,g_3$ 
to SM gauge bosons $\gamma,W^\pm, Z^0$ and gluon. The spectrum 
of these massive composite bosons and fermions
preserves the SM chiral gauge symmetries. The gauge and mass
eigenstates of massive composite bosons and fermions are the same.  
The interacting 1PI vertexes of composite particles 
and SM gauge bosons can be obtained by Ward identities associating 
to SM gauge symmetries \footnote{Calculations can be found 
in Refs.~\cite{xue2000,xue2000prd} and their appendixes 
for the simple case of the 
four-fermion operator (\ref{4foa}), where composite particles carry the definite quantum numbers of the $SU_L(2)$-gauge symmetry, and 
their tree-level gauge-couplings to the $SU_L(2)$-gauge bosons can be obtained by the Ward identity associating to the $SU_L(2)$-gauge 
symmetry in the UV-domain of the symmetric phase.}. As a result, at tree-level 
we obtain the following effective Lagrangian for 
the first-family composite particles 
(\ref{boundb},\ref{boundd},\ref{boundl},\ref{boundem}). 

For the massive composite Dirac fermions $\Psi^{i}_D$ 
and $\Psi_S$ (\ref{boundb}-\ref{boundm}) the $u_R$-quark channel in the 
quark sector, the effective Lagrangian reads  
\begin{eqnarray}
{\mathcal L}&=& \sum_i \bar\Psi^{i}_D(i\gamma^\mu\partial_\mu - M^i_D)\Psi^{i}_D 
+\bar\Psi_S(i\gamma^\mu\partial_\mu - M_S)\Psi_S\nonumber\\
&-&
e\sum_iQ_i(\bar\Psi^{i}_D\gamma^\mu\Psi^{i}_D) A_\mu-(2/3)e(\bar\Psi_S \gamma^\mu\Psi_S) A_\mu \nonumber\\
&-& \frac{g_2}{\sqrt{2}}\sum_i\bar\Psi^{i}_D\gamma^\mu(T^+W^+_\mu+T^-W^-_\mu)
\Psi^{i}_D \nonumber\\
&-&\frac{g_2}{2\cos\theta_W}\left\{\sum_i\bar\Psi^{i}_D\gamma^\mu(g_V^i+g_A^i)\Psi^{i}_D
+\bar\Psi_S\gamma^\mu(g_V^1-g_A^1)\Psi_S\right\}Z_\mu,
\label{comf}
\end{eqnarray}
where $i$ remains as the $SU_L(2)$-isospin index.
For the $d_R$-quark channel, the composite particles made by $u_{Ra}\rightarrow d_{Ra}$, the coupling $g^1_{V,A}\rightarrow g^2_{V,A}$ 
and $(2/3)e\rightarrow (-1/3)e$. 
These massive composite fermions couple to the
gauge bosons $W^\pm_\mu = (W^1_\mu\mp i W^2_\mu)/\sqrt{2}$, 
$A_\mu= B_\mu\cos\theta_W
+ W^3_\mu\sin\theta_W$ and $Z_\mu= -B_\mu\sin\theta_W
+ W^3_\mu\cos\theta_W$. 
The Lagrangian (\ref{comf}) shows that the massive spectrum and interacting vertex are vector-like fully preserving the parity symmetry. 
Namely, each left-handed (right-handed) SM elementary fermion has its
counterpart of right-handed (left-handed) composite fermion with the same 
quantum numbers of the SM chiral gauge symmetries. They form a massive Dirac composite fermion vectorially coupling to the SM gauge bosons. 
 
\begin{figure}[t]
\begin{center}
\includegraphics[height=2.5in, width=1.0in, angle =270 ]{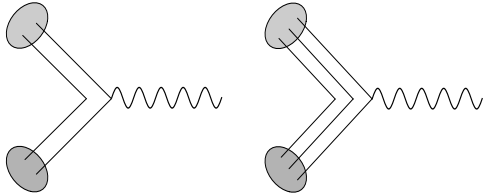}
\caption{We sketch the vertexes of composite boson (left) 
and fermion (right) interacting with SM gauge bosons (wave line), 
and the SM elementary fermion is represented by a solid line. In general, 
these vertexes should be understood as one-particle-irreducible vertexes of SM gauge bosons interacting with the entire composite particle as an elementary particle. 
At the lowest tree-level order of perturbative SM gauge couplings, 
the vertexes can be represented by the 
sum of possible gauge couplings to an SM elementary particle inside a 
composite particle.
Note that we do not show the vertexes of composite boson (left) interacting with two SM gauge bosons that are present in the effective Lagrangian (\ref{comc}).}\label{vertex}
\end{center}
\end{figure}

For the massive composite 
bosons $SU_L(2)$-doublet ${\mathcal A}^i$ (\ref{boundb}) 
for the $u_R$ channel, i.e., the 
neutral component ${\mathcal A}^1\sim (\bar u_{Ra} u^{a}_L)$ 
and charged component ${\mathcal A}^2
\sim (\bar u_{Ra} d^{a}_L)$ of the $U_Y(1)$-hypercharge $Y=-1/2$, the 
effective Lagrangian reads 
\begin{eqnarray}
{\mathcal L}&=& \sum_i\left[|{\mathcal D}{\mathcal A}^i|^2 + M^
2_\Pi|{\mathcal A}^i|^2\right],
\label{comc}
\end{eqnarray}
where
\begin{eqnarray}
{\mathcal D}{\mathcal A}^i &=& (\partial_\mu + ig_2{\bf\sigma}\cdot {\bf W_\mu }/2+ i g_1 Y B_\mu)
{\mathcal A}^i\nonumber\\
&=& \Big[\partial_\mu + i\frac{g_2}{\sqrt{2}}\left(T^+W^+_\mu+T^-W^-_\mu\right)+ ie \left(\matrix{\sin^{-1} 2\theta_WZ_\mu & 0\cr
0 &-\tan^{-1} 2\theta_W Z_\mu- A_\mu}\right)\Big]{\mathcal A}^i
\label{dcomcu}
\end{eqnarray}
and ${\bf\sigma}$ is the Pauli matrix in the isospin space. Instead, 
for the massive composite bosons $SU_L(2)$-doublet 
${\mathcal A}^i$ for the $d_R$ channel, i.e., the 
charged component ${\mathcal A}^1\sim (\bar d_{Ra} u^{a}_L)$ 
and neutral component ${\mathcal A}^2
\sim (\bar d_{Ra} d^{a}_L)$ of the $U_Y(1)$-hypercharge $Y=1/2$, 
Equation (\ref{dcomcu}) becomes
\begin{eqnarray}
{\mathcal D}{\mathcal A}^i &=& (\partial_\mu + ig_2{\bf\sigma}\cdot {\bf W_\mu }/2+ i g_1 Y B_\mu)
{\mathcal A}^i\nonumber\\
&=& \Big[\partial_\mu + i\frac{g_2}{\sqrt{2}}\left(T^+W^+_\mu+T^-W^-_\mu\right) + ie \left(\matrix{\tan^{-1} 2\theta_W Z_\mu + A_\mu & 0\cr
0 & -\sin^{-1} 2\theta_WZ_\mu}\right)\Big]{\mathcal A}^i.
\label{dcomcd}
\end{eqnarray}
The last matrix-terms in Eqs.~(\ref{dcomcu}) and (\ref{dcomcd}) 
respectively show that the isospin components ${\mathcal A}^1\sim \bar u_R u_L$ and
${\mathcal A}^2\sim \bar d_R d_L$ have no electric charge, 
but the $U_Y(1)$ hypercharge coupling to $Z^0$. The gauge
couplings in the effective Lagrangian 
(\ref{comf}), (\ref{dcomcu}) and (\ref{dcomcd}) are consistent with 
SM gauge bosons couplings to the elementary fermions inside the composite particles, see Fig.~\ref{vertex}. 

The effective Lagrangian 
(\ref{comf}), (\ref{dcomcu}) and (\ref{dcomcd}) can be generalized to 
the lepton sector and lepton-quark sector. Note that in the effective Lagrangian 
(\ref{comf}) and (\ref{comc}), we only present  
the spectra (kinetic terms or propagators) of massive composite particles and their
tree-level gauge-coupling to the elementary SM gauge bosons. 

To the leading order of the expansion in the powers of 
the perturbative SM gauge couplings (the tree-level), 
the interacting vertexes of composite bosons and fermions 
coupling to SM gauge bosons ($\gamma, W^\pm,Z^0$) 
are represented by the tree-level Feynman diagrams, 
see Fig.~\ref{vertex},
which show gauge bosons interacting with one of elementary particles 
inside composite particles.
Therefore, this implies that at the tree level,  some 
calculations of Feynman diagrams are similar to those in the SM, 
however, the fermion spectra and coupling vertexes are vectorlike, 
see Eqs.~(\ref{smf}) and (\ref{comf}). The final results are
characterized by the mass scales $M_{\Pi,F}$ (\ref{boundm}) of 
composite fermions 
\begin{eqnarray}
M_F > M_\Pi\propto\E_\xi,
\label{uscale} 
\end{eqnarray}
in the UV domain and functions of the SM gauge couplings $g_1,g_2, g_3$ 
at this mass scale $M_{\Pi,F}$, 
rather than the electroweak scale $v\approx 239.5$ GeV 
in the IR domain.

Since the propagators 
of massive composite particles are obtained from 
the lowest nontrivial contribution of the strong-coupling expansion \cite{xue1997}, it cannot be precluded that there could be 
the interacting vertexes 
between massive composite particles, e.g., the Yukawa-type interactions 
of composite bosons and fermions, stemming from the high-order contributions of the strong-coupling expansion. 
In this article, 
we do not attempt to discuss the interacting vertexes 
among composite particles and their relevance in the UV-domain.

\subsubsection{Some discussions on FCNC and anomalies in UV-domain}\label{uvfcnc}

We attempt to have some discussions on the analogical FCNC processes of the UV effective Lagrangian (\ref{comf}) and (\ref{comc}) of composite particles in the UV-domain. The composite-fermion case is considered as an example for 
discussions that apply also to the composite-boson case.
Recall that (i) composite fermions $\Psi^f$ are composed by the SM elementary fermions $\psi^f$ in the same family ``$f$''; (ii) 
the fermion-family ``diagonalized'' effective Lagrangian 
(\ref{comf},\ref{comc}) of massive composite particles 
$\Psi^f$ is obtained from the fermion-family ``diagonalized'' 
four-fermion operators (\ref{q1})
in the mass eigenstates of the SM elementary fermions $\psi^f$; 
(iii) the mass eigenstates of composite fermions 
coincide with the mass eigenstates of the SM elementary fermions in the scenes that composite fermions are composed by elementary fermions 
in their mass eigenstates, see Sec.~\ref{composites};
(iv) the mass and gauge eigenstates of composite fermions are 
{\it identical}, due to the exact conservation of the SM chiral gauge symmetries.

In the representation of mass eigenstates of composite fermions, 
totally contrary to the IR effective Lagrangian for the SM in the IR domain, 
see Sec.~\ref{ird} and in particular Sec.~\ref{irfcnc}, 
the UV effective Lagrangian (\ref{comf}) and (\ref{comc}) are realized 
in the scaling UV-domain of the {\it symmetric phase}, 
where the expectational values of two-fermion operators (\ref{mqu})
$\langle\psi^{f}_{_R} \bar \psi^{f}_{_L}\rangle\propto m^f\equiv 0, 
f=1,2,3$, identically vanish. The strong-coupling dynamics that forms massive composite particles in the symmetric phase is totally independent of the SM elementary fermion masses $m_f$ in the symmetry-breaking phase. 
The SM chiral-gauge symmetries and 
global fermion-family symmetry are exactly preserved 
by the strong-coupling dynamics and massive spectrum of composite fermions. 
More precisely, in the representation of mass eigenstates of composite 
fermions, the fermion-family number associating to the symmetry 
$U^f_L(1)\times U^f_R(1)$ $f=1,2,3$ for the 
composite fermion $\Psi^f$ in each family is exactly conserved.    
Therefore, any process violating fermion-family 
number is impossible, namely any ${\mathcal FCNC}$ process is completely prohibited. Here the notation ${\mathcal FCNC}$ refers to the process of changing one 
{\it composite} fermion $\Psi^{f}$ flavor 
to another {\it composite} fermion $\Psi^{f'}$ flavor 
with the same electric charge, which is different from the notion FCNC 
of the IR effective Lagrangian for the SM elementary fermions 
in the IR-domain.

However, as perturbatively turning on chiral-gauge-boson 
interactions to composite 
fermions, does the interactions between the chiral-gauge-boson 
$W^\pm$ and composite fermions introduce a suppressed 1PI vertex for the ${\mathcal FCNC}$ at loop level, analogous to that of the SM effective Lagrangian in the IR domain? The answer is no, for the reason that the mass and gauge eigenstates of composite fermions are identical, and 
fermion-family mixing matrices (\ref{mmud}) do not associate to the interacting vertexes of the $W^\pm$ boson and composite 
fermions (\ref{comf}). In conclusion, 
fermion-family-number violating processes like ${\mathcal FCNC}$ are impossible for the effective Lagrangian (\ref{comf}) and (\ref{comc}) 
in the UV-domain. The remained question is how the suppressed 1PI vertexes 
of the FCNC from the SM $W^\pm$-boson loop and CKM-type mixing matrices 
in the IR-domain is induced, when composite fermions decay 
into SM elementary fermions, a phase transition occurs 
from symmetric phase ($m_f=0$) to symmetry-breaking phase ($m_f\not=0$). 
This is deserved to study in another article.

In the SM Lagrangian (\ref{smf}) of chiral-gauge symmetries and chiral-gauged fermion spectra, 
the chiral-gauge anomalies associating to the $SU_L(2)\times U_Y(1)$ chiral-gauge 
symmetries are canceled, because of the $SU_L(2)$ Lie algebra and the 
SM fermion content of quarks and leptons in each fermion family, and the non conservation
of fermion number ($B-L$) is given by the instanton contribution. 
In the effective Lagrangian (\ref{comf}) of composite particles, 
although the gauge symmetries are still chiral, i.e., $SU_L(2)\times U_Y(1)$, the
spectra of composite fermions $\Psi_D$ and $\Psi_S$ are vector-likely gauged. As a result, 
the chiral-gauge anomalies from the left-handed composite fermions are exactly canceled by the anomalies from their right-handed counterparts. 
In the effective Lagrangian (\ref{comf}) of composite fermions, we obtained \cite{xue2000} the non conservation of fermion number ($B-L$) by considering the mixing anomaly \cite{preskill1991}.

As the running energy scale $\mu$ decreasing, and composite particle's 
form factor and binding energy vanishing, the 
composite particles become unstable and dissolve (decay) to 
their constitutes of SM elementary particles as final states.
In the following, we discuss the decay and annihilation channels of 
the composite particles into the SM elementary particles, 
in particular two gauge bosons, as final states.

\section{\bf Composite pesudo scalar bosons decay}\label{cbd}

This section \ref{cbd} and next section \ref{cfc} discuss the decays and other relevant processes of composite bosons and fermions 
interacting with SM gauge bosons described by the effective 
Lagrangians (\ref{comf}) and (\ref{comc}) of the effective theory 
of composite particles in UV-domain. These results provide possible 
connections with the diboson and other channels 
in ATLAS and CMS experiments.    

\subsection{Low-energy QCD pions and their decays}

First let us recall the low-energy 
hadronic physics of mesons (baryons) that 
are the bound states of two quarks (three quarks) due to the 
confinement dynamics of the local $SU(3)$-color gauge theory (QCD). 
This can be approximately described by 
the effective low-energy Lagrangian of baryons and mesons, which 
could be realized in the possible low-energy IR-domain of the QCD theory. 
On the other hand, asymptotically-free states of quarks and gluons 
are realized in the high-energy UV-domain of the QCD theory. 
In both bound states and asymptotically-free states, elementary quarks have the same electroweak couplings to the SM gauge bosons 
$W^\pm$, $Z^0$ and $\gamma$. 

To effectively study the non-perturbative nature of low-energy QCD theory, 
the global chiral-symmetric Lagrangian ($\sigma$-model)  
was adopted to describe the massless 
QCD-bound states of elementary $u$ and $d$ quarks: proton $p(uud)$, neutron $n(ddu)$ and pions ${\bf \pi}^{0,\pm}$(uu,dd,ud).
The spontaneous breaking of the $SU_L(2)\times SU_R(2)$ chiral symmetries 
leads to the massive doublet baryon fields $(p,n)$ and massless triplet 
pion fields $\pi^k$ as three Goldstone bosons. The explicit breaking induced by the $u$ and $d$ quark masses leads to massive pion fields 
$\pi^k$ and the partial conservation of the axial current (PCAC) 
\begin{eqnarray}
\partial^\mu A^j_\mu(x) =m_\pi^2 f_\pi \pi^j(x)\label{pcac}
\end{eqnarray} 
where the quark doublet $\psi^i=(u,d)$ and the axial current 
$A^j_\mu(x)=\bar \psi\gamma_\mu\gamma_5(\sigma^j/2)\psi$. The latter 
couples to the pion fields $\pi^j(x)$ due to the spontaneous breaking of 
chiral symmetries, leading to the nontrivial matrix elements
between the pion state $|\pi^k(p)\rangle$ and the vacuum
\begin{eqnarray}
\langle 0| A^j_\mu(x) |\pi^k(p)\rangle = ip_\mu f_\pi\delta^{jk} e^{-ipx},
\quad \langle 0| \partial^\mu A^j_\mu(x) |\pi^k(p)\rangle = m^2_\pi f_\pi\delta^{jk} e^{-ipx}.
\label{pcac1}
\end{eqnarray} 
The first matrix element defines the pion decay constant (form factor) 
$f_\pi$, the second matrix element and the mass-shell condition $p^2=m_\pi^2$ defines the pion mass $m_\pi$.
The charged pions $\pi^\pm=(\pi^1\mp i\pi^2)/2=
\bar d \gamma_5 u, \bar u\gamma_5 d$ and 
neutral pion $\pi^0=\pi^3/2=(\bar u \gamma_5 u-\bar d \gamma_5 d)/2$. 
For the isospin $j,k=1,2$ components in Eq.~(\ref{pcac}), the first matrix element determines 
the rate of the decay $\pi^+ (\bar u d)\rightarrow \mu^+ +\nu_\mu$
\comment{, via the weak interaction of the gauge $W$-boson exchange,
\begin{eqnarray}
\Gamma_{\pi^+ \rightarrow \mu^+ +\nu_\mu} \approx \frac{G^2m_\mu^2f_\pi^2(m_\pi^2-m_\mu^2)^2}{4\pi m_\pi^2},
\label{pi+}
\end{eqnarray}
}
and experimental value $f_\pi\approx 93$ MeV. In addition, for the isospin $j=k=3$
component in Eq.~(\ref{pcac}) reads 
\begin{eqnarray}
\partial^\mu A^3_\mu =m_\pi^2 f_\pi \pi^0 -\frac{\alpha}{8\pi}~
\epsilon_{\mu\nu\rho\sigma}F^{\mu\nu}F^{\rho\sigma},
\label{pcaca} 
\end{eqnarray}
receiving an axial anomaly in terms of the gauge 
field (photon) strength $F$
and fine-structure constant $\alpha$, contributed from the triangle diagram shown in Fig.~\ref{triangle}. This axial anomaly dominates 
the $\pi^0$-decay rate,
\begin{eqnarray}
\Gamma_{\pi^0\rightarrow \gamma\gamma} &= &\left(\frac{\alpha N_c}{3\pi f_\pi}\right)^2
\frac{m^3_{\pi^0}}{64\pi},
\label{pi0}
\end{eqnarray}
in excellent agreement with the experimental value. 

\subsection{Scalar and pesudo scalar composite bosons}

\comment{Associating to the global $SU_L(2)\times U_Y(1)$ chiral symmetries of the Lagrangian, 
the massive composite boson ${\mathcal A}^i\sim (\bar u_R\psi^i_L)$ 
or ${\mathcal A}^i\sim (\bar d_R\psi^i_L)$ is 
the fundamental representation of 
$SU_L(2)\times U_Y(1)$ chiral symmetric group of the $U_Y(1)$-hypercharge $Y=-1/2$ or $Y=1/2$. }

We attempt to study 
the decays of composite bosons ${\mathcal A}^i$ in the UV-domain, 
discussed in Secs.~\ref{composites} and \ref{efflag}. 
For the sake of simplicity, we 
adopt the first quark family $\psi^i_L=(u,d)_L$ and $\psi_R=u_R,d_R$ for illustrations. For the $u$-channel, the massive composite boson 
${\mathcal A}^i=(\bar u_R\psi^i_L)$ (\ref{boundb}), which is an $SU_L(2)$ 
doublet and $U_Y(1)$ charged $Y=-1/2$, can expressed in terms of scalar 
and pseudo scalar fields,
\begin{eqnarray}
{\mathcal A}^i=(\bar u_R\psi^i_L)=(1/2)\left[(\bar u\psi^i)
-(\bar u\gamma_5\psi^i)\right].
\label{sacomp}
\end{eqnarray}
where the form factor $[Z^{^S}_\Pi]^{-1/2}$ is omitted for simplifying notations. There are four components    
\begin{eqnarray}
S^0&=&\bar uu,\quad S^{-} =\bar ud\,; \quad 
\quad \Pi^0=\bar u\gamma_5u, 
\quad \Pi^-=\bar u\gamma_5d 
\label{4comp}
\end{eqnarray} 
where the isospin index $i=1,2$ are relabeled as $i'=0,\pm$ 
indicating the neutral or charged component. 
For the $d$-channel, by the substitution $u\rightarrow d$ in 
Eqs.~(\ref{sacomp}) and (\ref{4comp}), we have
\begin{eqnarray}
{\mathcal A}^i&=&(\bar d_R\psi^i_L)=(1/2)\left[(\bar d\psi^i)
-(\bar d\gamma_5\psi^i)\right],\nonumber\\
S^0&=&\bar dd,\quad S^+ =\bar du\,; \quad 
\quad \Pi^0 =\bar d\gamma_5d, 
\quad \Pi^+=\bar d\gamma_5u.
\label{4compd}
\end{eqnarray}
The neutral state $\Pi^0$ has the contributions from Eqs.~(\ref{4comp}) and (\ref{4compd}) that can be mixed up, but is different from the normal pion $\pi^0=(\bar uu-\bar dd)/\sqrt{2}$ in the $SU(3)$ quark model where the minus sign comes from the isospin $\tau^3=(1,-1)$. 
The discussions can be 
generalized to other families including leptons. The second and third 
families of composite bosons are simply replicated by the replacements 
$u\rightarrow c, t$ and $d\rightarrow s, b$. There is no mixing among three
families of composite bosons. The gauge interactions 
at the leading order (tree-level) of gauge couplings (\ref{dcomcu}) 
and (\ref{dcomcd}) do not introduce the mixing of 
composite bosons (\ref{4comp}) with its counterparts 
in the second or third family.

These composite bosons $\Pi^{0,\pm}$ (\ref{4comp}) and (\ref{4compd}) 
are bound by the strong-coupling dynamics in the UV domain, 
differently from the QCD-bound mesons. They have the same 
{\it chiral-invariant} mass $M_\Pi$ (\ref{boundm}). 
Instead the three pion fields $\pi^{0,\pm}$ 
have the {\it chiral-variant} mass $m_\pi$.
The vectorial and axial currents associating to the global $SU_L(2)\times U_Y(1)$ chiral symmetries are exactly conserved because these chiral symmetries are preserved, apart from the chiral gauge anomaly (axial anomaly). Instead the axial current
associating to the global chiral symmetries of the effective low-energy 
Lagrangian of the QCD-bound states are partially conserved because of spontaneous and explicit chiral-symmetry-breaking (PCAC). 
However, the massive pseudo scalar fields $\Pi^{0,\pm}$ 
(\ref{4comp},\ref{4compd})  and 
the QCD pion fields $\pi^{0,\pm}$ have the same quantum numbers of the SM symmetries. Moreover, we shall show the pseudo-scalar fields $\Pi^{0,\pm}$
decay into SM elementray particles, in particular the $\Pi^{0}$-decay 
into two SM gauge bosons, analogously to the decay $\pi^0\rightarrow \gamma\gamma$, attributing to the axial anomaly (\ref{pi0}).

\subsection{ Composite particle $\Pi^{0}$ decay into two SM gauge bosons}

\subsubsection{ $\Pi^{0}$ decay into two photons}

Suppose the vacuum state is $|{\bf 0}\rangle$ in the UV domain, the
composite-boson momentum state $|{\mathcal A}^{i}(q)\rangle$ on the mass 
shell can be defined by the field operator 
${\mathcal A}^{i}(q)$ or ${\mathcal A}^{i}(x)$ as follow
\begin{eqnarray}
|{\mathcal A}^{i}(q)\rangle \equiv
\frac{1}{F_\Pi} {\mathcal A}^{i\dagger}(q) | {\bf 0}\rangle = \frac{1}{F_\Pi} 
\int d^4x e^{-iqx} {\mathcal A}^{i\dagger}(x) | {\bf 0}\rangle ,
\label{cstate1}
\end{eqnarray}
with the normalization
\begin{eqnarray}
\langle {\mathcal A}^{i}(q)|{\mathcal A}^{i'}(q')\rangle = 
\delta^{ii'}(2\pi)^4\delta^4(q-q')\label{cstate1'},
\end{eqnarray}
or equivalently 
\begin{eqnarray}
\langle {\bf 0}| {\mathcal A}^i(x) |{\mathcal A}^{i'}(q)\rangle = F_\Pi\delta^{ii'} e^{iqx},
\label{cstate2}
\end{eqnarray}
where $F^{-1}_\Pi=F^{-1}_\Pi(q^2)$ relates to the form factor 
$[Z^{^S}_\Pi]^{-1/2}$ (\ref{boundb}), as discussed 
in Sec.~\ref{composite}. As will be seen soon, $F_\Pi$ 
is in fact the composite-boson decay constant. 
From Eq.~(\ref{cstate2}), 
the matrix element reads
\begin{eqnarray}
\langle {\bf 0}| \partial^\mu{\mathcal A}^i(x) |{\mathcal A}^{i'}(q)\rangle = iq^\mu F_\Pi\delta^{ii'} 
e^{iqx}.
\label{cstate4}
\end{eqnarray} 
The effective Lagrangian 
(\ref{comc}) gives the equation of motion 
and the mass-shell condition $q^2=M_\Pi^2$ of the composite 
boson ${\mathcal A}^i(x)$. From Eq.~(\ref{cstate4}), we have 
\begin{eqnarray}
\langle {\bf 0}|(\partial^\mu)^2{\mathcal A}^i(x) |{\mathcal A}^{i'}(q)\rangle
= M^2_\Pi F_\Pi\delta^{ii'} 
e^{iqx}.
\label{cstate5}
\end{eqnarray}
The matrix element (\ref{cstate2}) relates to the composite-boson propagator,
\begin{eqnarray}
\langle {\bf 0}| {\mathcal A}^i(0) |{\mathcal A}^{i'}(q)\rangle 
= \frac{1}{F_\Pi}\int d^4x \langle {\bf 0}| {\mathcal A}^i(0) {\mathcal A}^{i'\dagger}(x)|{\bf 0}\rangle e^{-iqx}
=\frac{\delta^{ii'}F^{-1}_\Pi(q^2)}{q^2-M^2_\Pi} .
\label{cstate2'}
\end{eqnarray} 
The matrix element of composite particle $\Pi^0$ in Eqs.~(\ref{4comp}) and (\ref{4compd}) reads, 
\begin{eqnarray}
\langle f(p')| \Pi^0(0) |f(p)\rangle 
=\frac{g_{_{\Pi^0}}}{-q^2+M^2_\Pi} \bar u_f(p') \gamma_5 u_f(p),\quad q=p'-p
\label{cstate2''}
\end{eqnarray}
where the $g_{_{\Pi^0}}=g_{_{\Pi^0}}(q^2)$ is the $\Pi^0$-coupling 
to its constituent fermions $f=u,d$ 
quarks of mass $m_f$ and wave function $u_f$. Actually, the coupling
$g_{_{\Pi^0}}(q^2)$ is the 1PI
vertex function of composite particle $\Pi^0$ and its constituent fermions and we parametrize it by the 
form factor $F^{-1}_\Pi(q^2)$ and constituent mass $m_f$   
\begin{eqnarray}
g_{_{\Pi^0}}(q^2)=
m_fF^{-1}_\Pi(q^2),
\label{gf}
\end{eqnarray}
and $g_{_{\Pi^0}}$ is finite as $m_f\rightarrow 0$, 
see for example \cite{peskin_book}. 
These definitions and equations apply also for scalar composite bosons 
$S^i$ and pseudo scalar composite bosons $\Pi^i$, see Eqs.~(\ref{4comp}) 
and (\ref{4compd}).

\begin{figure}[t]
\begin{center}
\includegraphics[ ]{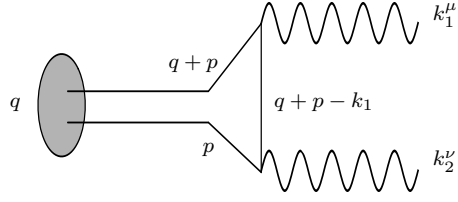}
\caption{We show the triangle diagram (a quark loop of solid line) of a composite boson (two parallel solid lines associating to a pseudo-scalar vertex $\gamma_5$) 
decaying into the two SM gauge bosons (two wave lines associating to 
two vector-like vertexes $\gamma_\mu$ and $\gamma_\nu$).} \label{triangle}
\end{center}
\end{figure} 

We are in the position of discussing the rate of the 
neutral $\Pi^0$ decay into two SM gauge bosons, i.e., diboson channels:
\begin{eqnarray}
\Pi^0\rightarrow G+G^\prime,
\label{a0}
\end{eqnarray} 
where $G$ and $G^\prime$ represent the SM gauge bosons 
$W^\pm, Z^0$ and $\gamma$. The amplitude of the 
decay (\ref{a0}) is defined by the matrix element
\begin{eqnarray}
{\mathcal F}(q^2)\Big|_{q^2\rightarrow M^2_{\Pi^0}} 
=\epsilon_1^\mu(k_1)\epsilon_2^\nu(k_2) 
T_{\mu\nu}(k_1,k_2,q)\Big|_{q^2\rightarrow M^2_{\Pi^0}},\quad q=k_1+k_2
\label{a1}
\end{eqnarray}
and 
\begin{eqnarray}
T_{\mu\nu}(k_1,k_2,q) &=& (q^2-M^2_{\Pi^0})(gg')
\int d^4x d^4y e^{ik_1x+ik_2y} \langle {\bf 0}|T(
J_\mu(z)J'_\nu(y) \Pi^{0}(0))|{\bf 0}\rangle ,
\label{a3}
\end{eqnarray}
where $k_{1,2}$ and $\epsilon_{1,2}$ are four momenta and polarizations of the SM gauge bosons $G$ and $G^\prime$, and $J_\mu(z)$ [$J'_\nu(y)$] is fermion current coupling $g~ [g']$
to the SM gauge boson $G ~[G^\prime]$. 
The amplitude $T_{\mu\nu}(k_1,k_2,q)$ has the symmetry $(k_1,\mu)\leftrightarrow
(k_2,\nu)$.  
\comment{ The vertex function in Eq.~(\ref{a1}) reads   
\begin{eqnarray}
\Gamma_{\mu\nu}(k_1,k_2,q) &=& gg'\int d^4zd^4y e^{ik_1z+ik_2y} \langle {\bf 0}|T(
J_\mu(z)J'_\nu(y) )|\Pi^{0}(q)\rangle ,
\label{a2}
\end{eqnarray}
Using Eqs.~(\ref{cstate1}-\ref{cstate2'}), see the standard reduction formula Bjorken and Drell, the vertex function can be written as
\begin{eqnarray}
\Gamma_{\mu\nu}(k_1,k_2,q) &=& (q^2-M^2_\Pi)\frac{gg'}{F_\Pi}\int d^4x d^4yd^4z e^{ik_1z+ik_2y-iqx} \langle {\bf 0}|T(
J_\mu(z)J'_\nu(y) \Pi^{0}(x))|{\bf 0}\rangle ,
\label{a3}
\end{eqnarray} 
where the pseudo scalar composite field 
$\Pi^{0}(x)=\bar u\gamma_5 u, \bar d\gamma_5 d$ 
in Eqs.~(\ref{4comp}) or (\ref{4compd}).  
The vertex (\ref{a3}) has the Lorentz covariant structure,
\begin{eqnarray}
\Gamma_{\mu\nu}(k_1,k_2,q) &=& (q^2-M^2_\Pi)\frac{(gg')}{F_\Pi}\epsilon_{\mu\nu\rho\sigma}k_1^\rho k_2^\sigma \left[\Gamma_\Pi^{0}(q^2,m^2)+\frac{\Gamma(q^2,m^2)}{(q^2-M^2_\Pi)}\right] ,
\label{a3'}
\end{eqnarray}
where the first finite term $\Gamma_\Pi^{0}(q^2,m^2)$ represents 
the $\Pi^{0}$-coupling vertex to an elementary fermion of mass $m$, 
the second term has the pole $q^2=M^2_\Pi$ and $\Gamma(q^2,m^2)$ relates to the 
form factor of the process (\ref{a0}). In the limit of $q^2\rightarrow M^2_\Pi$, 
the $\Pi^{0}$-decay amplitude (\ref{a1}) on the mass shell, we have
\begin{eqnarray}
\Gamma_{\mu\nu}(k_1,k_2,q)\Big|_{q^2\rightarrow M^2_\Pi} &=& \frac{(gg')}{F_\Pi}\epsilon_{\mu\nu\rho\sigma}k_1^\rho k_2^\sigma  ~\Gamma(q^2,m^2).
\label{a3''}
\end{eqnarray}  
}

The lowest one-loop contribution to
the vertex function (\ref{a3}) is represented by the triangle 
Feynman diagram, see Fig.~\ref{triangle}. Considering 
Eqs.~(\ref{cstate2''}) and (\ref{gf}), the $\Pi^0$-decay amplitude $T_{\mu\nu}$ 
(\ref{a3}) computed at this one-loop level is given by
\begin{eqnarray}
T_{\mu\nu}(k_1,k_2,q)&=&T^{(1)}_{\mu\nu}(k_1,k_2,q,m_f)+T^{(2)}_{\nu\mu}(k_2,k_1,q,m_f),
\label{loop0}\\
T^{(1)}_{\mu\nu}(k_1,k_2,q,m_f) &=& -(gg')N_cm_fF^{-1}_\Pi(q^2)\times\nonumber\\
&\times&\int\frac{d^4p}{(2\pi)^4}
\frac{\tr[({\slashed q}+{\slashed p}-{\slashed k_1}+m_f)
\Gamma_\mu({\slashed q}+{\slashed p}+m_f)\gamma_5({\slashed p}+m_f)\Gamma'_\nu]}{[(q+p-k_1)^2-m_f^2][(q+p)^2-m_f^2](p^2-m_f^2)} ,
\label{loop}
\end{eqnarray}
where 
the coupling vertexes $g\Gamma_\mu$ and $g'\Gamma'_\nu$ 
to the SM gauge bosons $G$ and $G^\prime$ are given by 
the SM Lagrangian (\ref{comf}), and the trace ``$\tr$'' is over the spinor space. Equation (\ref{loop}) is not well-defined 
because of the linear divergence of the momentum integral. 
Introducing the Pauli-Villars mass ${\mathcal M}$ 
that plays the role of the UV 
cutoff $\Lambda$,  we adopt the Pauli-Villars regularization,
\begin{eqnarray}
T^{(1)}_{\mu\nu}(k_1,k_2,q,m_f) \rightarrow T^{(1)}_{\mu\nu}(k_1,k_2,q,m_f)-T^{(1)}_{\mu\nu}(k_1,k_2,q,{\mathcal M}),
\label{reg}
\end{eqnarray}
to make the momentum integral to be finite and well-defined. Note that the Pauli-Villas regularization ${\mathcal M}$ explicitly breaks chiral gauge symmetries of effective Lagrangian. 

For the process $\Pi^0\rightarrow \gamma+ \gamma$, $g=g'=e~Q^i$,
$\Gamma_\mu=\gamma_\mu$  and $\Gamma'_\nu=\gamma_\nu$ 
in Eq.~(\ref{loop}). The $J_\mu(z)$  and $J'_\nu(y)$ in Eq.~(\ref{a3}) 
are vector-like and conserved fermion currents coupling with two on-shell photons $k_1$ and $k_2$. 
The trace in the numerator of Eq.~(\ref{loop}) is equal to 
$4im_f \epsilon_{\mu\nu\rho\sigma}k_1^\sigma k_2^\rho$.
With the on-shell conditions
$k_1^2= 0$, $k_2^2=0$ and $2k_1\cdot k_2= q^2$, 
the two contributions (\ref{loop0}) to the amplitude (\ref{a1}) 
of $\Pi^0$ decaying into two photons is given by
\begin{eqnarray}
{\mathcal F}(q^2)\Big|_{q^2\rightarrow M^2_{\Pi^0}} &=& \frac{(gg')N_c}{2\pi^2F_\Pi}\epsilon_{\mu\nu\rho\sigma} \epsilon^\mu_1 \epsilon^\nu_2 k^\rho_1 k^\sigma_2\times\nonumber\\
&\times&\int_0^1dx\int_0^{1-x}dy\left[\frac{{\mathcal M}^2}{{\mathcal M}^2-xy q^2}-\frac{m_f^2}{m_f^2-xy q^2}\right]_{q^2\rightarrow M^2_{\Pi^0}}
\label{a4}\\
&=&\frac{(gg')N_c}{4\pi^2F_\Pi}\epsilon_{\mu\nu\rho\sigma} \epsilon^\mu_1 \epsilon^\nu_2 k^\rho_1 k^\sigma_2.
\label{a4'}
\end{eqnarray} 
The second term in Eq.~(\ref{a4}) is suppressed for $M_{\Pi^0}\gg m_f$
and the first term for ${\mathcal M}\rightarrow \infty$ leading 
to the final result (\ref{a4'}). The result is just the axial anomaly of the triangle Feynman diagram (Fig.~\ref{triangle}) with one axial 
vertex ``$\gamma_5$'' and two vector-like (AVV) vertexes 
``$\gamma_5\gamma_\mu\gamma_\nu$''. This axial-anomaly 
result (\ref{a4'}) is due to the regularization 
that does not preserve chiral gauge symmetries of effective Lagrangian, and it is independent of high-order contributions, constituent fermion mass $m_f$ and UV cutoff ${\mathcal M}$. 
Equation (\ref{a4'}) shows that the $F_\Pi$ value at $q^2=M^2_{\Pi^0}$ 
is the $\Pi^0$-decay constant. For the first quark family, 
there are two $\Pi^0$ channels (\ref{4comp}) 
and (\ref{4compd}) with $Q_i=(2/3,-1/3)$, i.e., $g^2=e^2(4/9,1/9)$.
The total amplitude of $\Pi^0\rightarrow \gamma+ \gamma$ reads,
\begin{eqnarray}
{\mathcal F}(q^2)\Big|^{\Pi^0\rightarrow \gamma+ \gamma}_{q^2\rightarrow M^2_{\Pi^0}}
&=&\sum_i(Q^i)^2\frac{e^2N_c}{4\pi^2F_\Pi}\epsilon_{\mu\nu\rho\sigma} \epsilon^\mu_1 \epsilon^\nu_2 k^\rho_1 k^\sigma_2=\left(\frac{5}{9}\right){\mathcal F}\label{a4gg}\\
{\mathcal F}&\equiv&\frac{e^2N_c}{4\pi^2F_\Pi}\epsilon_{\mu\nu\rho\sigma} \epsilon^\mu_1 \epsilon^\nu_2 k^\rho_1 k^\sigma_2.
\label{def}
\end{eqnarray}    
The amplitude (\ref{a4gg}) should be multiplied by a factor $N_c=3$, 
if we consider three quark families. 

It needs some more clarifications on the 
amplitude (\ref{a4gg}). Both the composite pseudo-scalar boson $\Pi^0$ 
and the composite scalar Higgs boson $H$ decay into the two-photon state, however these two channels are not mixed. They have different parities, masses (invariant masses) and decay constants. The former composed by 
the strong-coupling dynamics in the UV-domain decay into two photons
via the pseudo-scalar coupling (see Fig.~\ref{triangle}) 
to the loop of $u$ and $d$ quarks. 
Whereas the latter composed by the NJL-dynamics 
in the IR-domain via the scalar coupling $\bar tt H$ (\ref{smf}) 
to a top-quark loop. 
In the LHC $pp$ collision, the 
first family of composite particles made of $u$ and $d$ quarks
are more probably produced, compared with other families of composite particles that are produced by small SM gauge interacting vertexes.

\subsubsection{ $\Pi^{0}$ decay into $\gamma+ Z^0, W^++ W^-$ and $Z^0+ Z^0$}

Let us study 
other possible decay channels $\Pi^0\rightarrow \gamma+ Z^0$, 
$\Pi^0\rightarrow W^++ W^-$ and $\Pi^0\rightarrow Z^0+ Z^0$. 
Suppose that the pseudo scalar composite field 
$\Pi^{0}(x)$ is heavy and its mass $M_{\Pi^0}$ is 
much larger than the $W^\pm$ and $Z^0$ masses, the intermediate gauge bosons $W^\pm$ and $Z^0$ are considered to be
approximately massless, $k_1^2\approx 0$, $k_2^2\approx 0$ and $2k_1\cdot k_2\approx q^2$. In addition, the coupling-vertexes $g\Gamma_\mu$ and $g' \Gamma'_\nu$ in Eq.~(\ref{loop}) are given in the SM Lagrangian (\ref{smf}), they contain both axial and vector-like vertexes. 
We consider the vector-like vertexes so that
the triangle Feynman diagram (Fig.~\ref{triangle}) and the amplitude 
(\ref{a3}) has the AVV structure, whose nontrivial axial-anomaly amplitude can be compared with the amplitude (\ref{a4gg}) of the channel 
$\Pi^0\rightarrow \gamma+ \gamma$. 
As a result, for the process 
$\Pi^0\rightarrow \gamma+ Z^0$, the couplings $g=eQ^i$ 
and $g'=g_2g_V^i/(2\cos\theta_W)$, we obtain from Eq.~(\ref{a4'})
the amplitude, 
\begin{eqnarray}
{\mathcal F}(q^2)\Big|^{\Pi^0\rightarrow \gamma+ Z^0}_{q^2\rightarrow M^2_{\Pi^0}}
&=&\frac{1}{\sin 2\theta_W}\sum_i(Q^ig_V^i){\mathcal F}\nonumber\\
&=&\frac{1}{\sin 2\theta_W}\left(\frac{1}{2}-\frac{5}{9}\sin\theta_W^2\right){\mathcal F}.
\label{a4gz}
\end{eqnarray}
For the process 
$\Pi^0\rightarrow W^++ W^-$, the couplings $g=g'=g_2/(2\sqrt{2})$ and we obtain from Eq.~(\ref{a4'})
the amplitude, 
\begin{eqnarray}
{\mathcal F}(q^2)\Big|^{\Pi^0\rightarrow W^++ W^-}_{q^2\rightarrow M^2_{\Pi^0}}
&=&\frac{1}{8\sin^2 \theta_W}\sum_i{\mathcal F}
=\frac{1}{8\sin^2 \theta_W}~{\mathcal F},
\label{a4ww}
\end{eqnarray}
where the sum over isospin ``$i$'' gives one in this case, see Eq.~(\ref{smf}).
For the process 
$\Pi^0\rightarrow Z^0+ Z^0$, the couplings $g=g'=g_2g_V^i/(2\cos\theta_W)$ and we obtain from Eq.~(\ref{a4'})
the amplitude, 
\begin{eqnarray}
{\mathcal F}(q^2)\Big|^{\Pi^0\rightarrow Z^0+ Z^0}_{q^2\rightarrow M^2_{\Pi^0}}
&=&\frac{1}{\sin^2 2\theta_W}\sum_i(g_V^i)^2{\mathcal F}\nonumber\\
&=&\frac{1/2-\sin^2\theta_W+(5/9)\sin^4\theta_W}{\sin^2 2\theta_W}~{\mathcal F}.
\label{a4zz}
\end{eqnarray}
Their differences from the two-photon amplitude (\ref{a4gg}) are 
only attributed to the different eletroweak couplings 
of the SM (\ref{smf}).

\subsection{Ratios of different channels of $\Pi^{0}$ decay 
into two SM gauge bosons}

The numerical rate of the neutral $\Pi^0(q)$ decay (\ref{a0})
at the mass shell ($q^2=M_{\Pi^0}^2$) is given by integrating 
the squared amplitude (\ref{a1}) over the phase 
space of two gauge bosons, 
\begin{eqnarray}
\Gamma_{\Pi^0\rightarrow G+G^\prime} &= & \frac{1}{2 M_{\Pi^0}}\sum_{\epsilon_1,\epsilon_2}
\int \frac{d^3k_1d^3k_2}{4(2\pi)^6\omega_1\omega_2}\Big|{\mathcal F}(q^2)\Big|^2_{{\Pi^0\rightarrow G+G^\prime}} \nonumber\\
&\times&(2\pi)^4\delta^4(q-k_1-k_2)\Big|_{q^2=M_\Pi^2}\nonumber\\
&= & {\mathcal C}_{\Pi^0\rightarrow G+G^\prime} ~ \Gamma,
\label{prate}\\
\Gamma &\equiv&\left(\frac{\alpha N_c}{3\pi F_\Pi}\right)^2
\frac{M^3_{\Pi^0}}{64\pi},
\label{deg}
\end{eqnarray}
where the mass $M_{\Pi^0}$ and decay constant $F_\Pi$ 
of the composite boson $\Pi^0$ are totally unknown, while 
the coefficients ${\mathcal C}_{\Pi^0\rightarrow G+G^\prime}$ are completely determined from Eqs.~(\ref{a4gg}), (\ref{a4gz}), (\ref{a4ww}) 
and (\ref{a4zz}) in terms of SM gauge couplings. 
Note that the decay rates (\ref{prate}) depend on the $\Pi^{0}$ 
mass $M_{\Pi^0}$ and decay constant $F_{\Pi}$ determined by 
the strong-coupling dynamics, analogously to the pion mass 
$m_{\pi}$ and decay constant $f_{\pi}$ determined by the QCD dynamics. 

\comment{
In Eqs.~(\ref{4comp}) and (\ref{4compd}), the pseudo-scalars form 
composite meson states $\Pi^0$ and $\Pi^\pm$ carry the same quantum numbers of QCD pions
$\pi^0$ and $\pi^\pm$ in the SM, however different masses $M_{\Pi^{0,\pm}}\gg m_{\pi^{0,\pm}}$. 
Apart from the different kinematic threshold, 
these composite bosons $\Pi^{0,\pm}$ undergo all decay channels of QCD pions.}

\comment{In the Feynman diagram representation at the tree-level, 
see Fig.~\ref{triangle},
these are the triangle diagram of 
quark loop with three interaction vertexes, one associates to the axial one
representing the composite boson $\Pi^0$ and other two are electroweak interactions associate to the SM gauge bosons.
These Feynman diagrams are the most relevant leading order contribution
to the composite boson decay rate.
} 
In the rest frame of the composite meson $\Pi^{0}$, 
the final states of two body decay follow the energy-momentum conservations 
${\bf p}_{1}=-{\bf p}_{2}, |{\bf p}_{1}|=|{\bf p}_{2}|$ 
and have the invariant mass ${\mathcal M}_{\rm inv}=M_{\Pi^{0}}$. 
The rates of the composite meson $\Pi^{0}$ decaying into two SM gauge bosons $G$ and $G^\prime$ are given by Eq.~(\ref{prate}), we then obtain the rates of diboson channel decay (\ref{a0}):
\begin{eqnarray}
\Gamma_{\Pi^0\rightarrow \gamma+\gamma} &= &\left(\frac{5}{9}\right)^2\Gamma,\label{2g}\\
\Gamma_{\Pi^0\rightarrow \gamma +Z^0} &= &\frac{1}{\sin^2 2\theta_W}\left(\frac{1}{2}-\frac{5}{9}\sin^2\theta_W\right)^2
\Gamma,\label{gz}\\ 
\Gamma_{\Pi^0\rightarrow W^++W^-}&= &\left(\frac{1}{8\sin^2 \theta_W}\right)^2\Gamma,
\label{ww}\\
\Gamma_{\Pi^0\rightarrow Z^0+Z^0}&= &\left(\frac{1/2-\sin^2\theta_W+(5/9)\sin^4\theta_W}{\sin^2 2\theta_W}\right)^2\Gamma.
\label{zz}
\end{eqnarray}
The diphoton channel (\ref{2g}) is expected to have the largest branching ratio, analogously to the QCD pion decay $\pi^0\rightarrow \gamma\gamma$. Suppose that the possible resonance 
of the diphoton channel (\ref{2g}) of the invariant mass 
${\mathcal M}_{\gamma\gamma}$ 
is observed in high-energy experiments,
it implies that the composite meson mass 
$M_{\Pi^{0}}\gtrsim {\mathcal M}_{\gamma\gamma}$, 
since there are other decay channels (\ref{gz},\ref{ww},\ref{zz}) into
two gauge bosons of $\gamma Z^0$, $W^+W^-$ and $Z^0Z^0$. The channel 
$\gamma Z^0$ corresponds to the final states of a photon and two leptons or 
a ``fat'' jet of two quarks. The channels 
$W^+W^-$ and $Z^0Z^0$ correspond to the final states of (i) 
two ``fat'' jets of four quarks for $W$ and $Z$ hadronic decays or 
(ii) four leptons for $W$ and $Z$ leptonic decay, or a ``fat'' jet 
and two leptons in the combination of cases (i) and (ii).

For the reasons that the properties of composite 
particles, e.g., form-factors and masses (or the binding-energy depth) 
due to the strong-coupling dynamics,
are unknown, we have not been able to calculate the total rate 
$\Gamma^{\rm total}_{\Pi^{0}}$ and width of the composite boson $\Pi^{0}$ 
that decays into the final states of the SM elementary particles including 
the channel of two gauge bosons or two 
fermions \footnote{In Ref.~\cite{xue2014}, 
we have some discussions of composite
boson decaying into the final state of two quarks.}. 
Nevertheless, at the leading order 
(tree-level) of gauge
interactions, we are able to calculate the rates
$\Gamma_{\Pi^{0}\rightarrow \gamma\gamma,\gamma Z,\cdot\cdot\cdot}$ 
(\ref{prate})
of the composite boson $\Pi^{0}$ decaying into two gauge bosons 
as diboson final states,
\begin{eqnarray}
\Pi^0&\rightarrow &\gamma \gamma, \gamma Z^0, W^+W^-, Z^0Z^0, \cdot\cdot\cdot.\label{pidecay}
\end{eqnarray} 
These diboson channels (\ref{pidecay}) 
are expected to be the most energetically favorable and 
largest branching ratio,
\begin{eqnarray}
B_{\Pi^{0}\rightarrow \gamma\gamma,\gamma Z^0,W^+W^-,Z^0Z^0}
= \frac{\Gamma_{\Pi^{0}\rightarrow \gamma\gamma,\gamma Z^0,W^+W^-,Z^0Z^0}}{\Gamma^{\rm total}_{\Pi^{0}}}.
\label{bratio1}
\end{eqnarray}
Moreover, in high-energy experiments, the diboson final state 
and its kinematics might be more easily 
identified than the final state of two fermions (quarks or two jets) 
due to the background of the QCD dynamics.

The total decay rate and width of the composite boson, 
as well as the branching ratios of decay channels, are very 
important for the collider phenomenology of the composite boson. 
However, the following ratios of the branching ratios (\ref{bratio1}) for 
different decay channels (\ref{pidecay}), ``relative branching ratio''
\begin{eqnarray}
\frac{\Gamma_{\Pi^{0}\rightarrow \gamma\gamma,\gamma Z^0,W^+W^-,Z^0Z^0}}
{\Gamma_{\Pi^{0}\rightarrow \gamma\gamma}}=\frac{B_{\Pi^{0}\rightarrow \gamma\gamma,\gamma Z^0,W^+W^-,Z^0Z^0}}
{B_{\Pi^{0}\rightarrow \gamma\gamma}}=\frac{\Gamma_{\Pi^{0}\rightarrow \gamma\gamma,\gamma Z^0,W^+W^-,Z^0Z^0}/\Gamma^{\rm total}_{\Pi^{0}}}
{\Gamma_{\Pi^{0}\rightarrow \gamma\gamma}/\Gamma^{\rm total}_{\Pi^{0}}},
\label{bratio}
\end{eqnarray} 
depend only on the SM gauge couplings $g$ and $g'$ at the 
energy scale $M_\Pi$, given by the effective Lagrangian (\ref{comf}). 
\comment{
These rates depend on the gauge 
couplings, see the effective Lagrangian (\ref{comf}),
which are actually 
the SM gauge couplings in Eq.~(\ref{smf}), since the SM gauge bosons
couple to the quark or lepton inside the composite particle.
Using the $W^\pm$-coupling $g_2/(2\sqrt{2})=e/(2\sqrt{2} \sin\theta_W)$ and the $Z^0$-coupling $\approx g_2/(4\cos\theta_W)$ to fermion fields,}
Following these discussions and using Eqs.~(\ref{2g}-\ref{zz}),
we can approximately estimate the decay-rate ratios (\ref{bratio})
\begin{eqnarray}
\Gamma_{\Pi^0\rightarrow \gamma Z^0} /\Gamma_{\Pi^0\rightarrow \gamma\gamma}&= & 
\frac{\left(9/10-\sin^2\theta_W\right)^2}{\sin^22\theta_W}\approx 0.63,\nonumber\\
\Gamma_{\Pi^0\rightarrow W^+W^-} /\Gamma_{\Pi^0\rightarrow \gamma\gamma}&\approx & 
\left(\frac{9}{40\sin^2 \theta_W}\right)^2\approx 0.96,\nonumber\\
\Gamma_{\Pi^0\rightarrow Z^0 Z^0} /\Gamma_{\Pi^0\rightarrow \gamma\gamma}&= &\left(\frac{(9/10)- (9/5)\sin^2\theta_W+\sin^4\theta_W}{\sin^2 2\theta_W}\right)^2\approx 0.58,
\label{br}
\end{eqnarray} 
where the value $\sin^2\theta_W\approx 0.23$ is approximately adopted to obtain numbers. These relations provide 
a possibility to verify the validity of such an effective theory of composite particles, if the resonance of diphoton channel and 
other diboson channels are also observed and measured in high-energy experiments,  
for instance the on-going ATLAS and CMS experiments 
in the LHC $pp$ collisions.

It is also possible that the composite boson 
$\Pi^0$ decays into the dijet final state of 
two quarks which was already discussed in Ref.~\cite{xue2014,xue2015}. 
Suppose that 
the neutral composite meson $\Pi^0$ is produced by $pp$ collisions 
at the LHC, its resonance location ${\mathcal M}_{\rm all~ channels}=M_{\Pi^0}$, 
and the rate of each decay channel depends on
the values of the mass $M_{\Pi^0}$, decay constant $F_\Pi$ 
and SM gauge couplings at these scales.  
Which channel 
is more relevant for detections, depending not only on its theoretical 
branching ratio in principle, 
but also on its experimental measurement in practice. 

The same analysis and discussion can be generalized 
to the decay rates of neutral 
composite bosons $\Pi^0$ made of 
charged lepton and/or neutrino pair (\ref{boundl}) 
in the lepton channel. Their decay constants and masses
should be approximately at the same scale of the $\Pi^0$ decay constant 
$F_{\Pi}$ and mass $M_{\Pi^0}$. Their decay rates can be obtained 
by Eqs.~(\ref{a4gg}-\ref{deg}) for $N_c=1$, couplings $Q^i$ and $g_V^i$ of the lepton sector in the SM Lagrangian (\ref{smf}). As an example 
for the two-photon final state,
the rate should be $(5/9)^2N_c^2=25/9$ time smaller than the 
rate (\ref{2g}) of the quark channel. 
However, the ratios of branching ratios for different diboson channels 
are similar to Eq.~(\ref{br}), but modified accordingly to
the leptonic $Q^i$ and $g_V^i$ values. 
Beside, in the LHC $pp$ collision, 
the production probability of leptonic composite bosons is smaller than 
hadronic one, due to the leptonic 
production rate is proportional to 
the small fine-structure constant $\alpha$.                 

\subsection{ $\Pi^{\pm}$ and other composite boson decays}

Equations (\ref{pcac}) and (\ref{pcac1}) of the PCAC are not applicable for the charged composite bosons $\Pi^{\pm}$ decay. Therefore,
we cannot use the analogy of the QCD charged 
pion $\pi^{\pm}$ decays to obtain the rates of $\Pi^{\pm}$ decay into either 
two quarks (jets) or two leptons,
\begin{eqnarray}
\Pi^+&\rightarrow& 
t + b, \cdot\cdot\cdot ; \tau^+ + \nu_\tau,
\cdot\cdot\cdot,
\label{pidecay+}
\end{eqnarray} 
and charge-conjugated processes. In the quark case, 
the composite bosons $\Pi^{\pm}$ are made of quarks (\ref{boundb}), 
and in the lepton case the composite 
bosons $\Pi^{\pm}$ are made of leptons (\ref{boundl}).
\comment{
In the charged channels of final states of leptons or quarks, we have 
\begin{eqnarray}
\Gamma_{\Pi^+\rightarrow \tau^++\nu_\tau
} &\approx &\frac{G_F^2}{4\pi}f^2_\Pi m_{\tau}^2
M_{\Pi^+} 
,\quad \cdot\cdot\cdot ,
\label{pidecayr1}\\
\Gamma_{\Pi^+\rightarrow t+b 
} &\approx &\frac{G_F^2}{4\pi}f^2_\Pi m_{t}^2 M_{\Pi^+}, 
\quad \cdot\cdot\cdot.
\label{pidecayr}
\end{eqnarray}
}
Nevertheless, it is expected that the $\Pi^{\pm}$-decay channel to the heaviest fermions,
top quark or tau lepton, should be most favorable.
Suppose that the charged composite meson $\Pi^\pm$ is produced by $pp$ collisions 
at the LHC, its resonance locates at the invariant mass
${\mathcal M}_{\rm qq}\approx M_{\Pi^\pm}$ of the dijet final state 
or other possible final states produced by two quarks (two jets). 
Whereas the charged lepton channel implies the final state of $\tau^\pm$ lepton of
energy $\sim M_{\Pi^{\pm}}/2$ and missing energy carried away by $\nu_\tau$-neutrino.

In Eqs.~(\ref{4comp}) and (\ref{4compd}), the scalar bosons form composite 
quarkonium states $\bar uu$ and $\bar dd$ that carry the same quantum 
numbers of QCD quarknium states, however, have much larger masses 
$\sim M_{\Pi^{0}}$. It is expected that apart from different kinematic 
threshold, these composite quarkonium states 
in principle undergo all decay channels of QCD quarkonium states. 
Recall that the direct decays of composite bosons $\Pi^\pm$ and quarkonium states into dijets or dilepton without an intermediate 
state were preliminarily discussed \cite{xue2014,xue2015}.  
On the basis of the SM chiral gauge symmetries and 
vector-like composite fermion 
content, there are also other possible final decay channels, analogously to those studied in the effective theory of QCD at low energies. 
These will be issues in future studies.

\section{\bf Composite fermion decay and annihilation channels}\label{cfc}

In this section \ref{cfc}, we turn to discussions of decay and annihilation channels of composite fermions, see Sec.~\ref{composites}, on the basis of the effective Lagrangian (\ref{comf}) of 
their kinetic terms and interactions to SM gauge bosons. We show the 
peculiar characteristics of composite fermions decay. It could be a possibility and possible criterion for the experimental verification or 
falsification of the effective theory of composite particles in the UV-domain of strong four-fermion couplings. 

\subsection{Composite fermions decay}\label{cfd}

\subsubsection{Decay into two gauge bosons and a quark}\label{cfdq}

For the sake of simplicity, we 
adopt the first quark family (\ref{boundb}-\ref{boundd}) to
discuss the decay of composite Dirac fermions, and 
discussions can be generalized to other families including leptons. 
The composite Dirac fermions $\Psi_D$ (\ref{boundd}) should be  
more massive than their constituents of composite boson ${\mathcal A}^i$ (\ref{boundb}), i.e., $M_F > M^{0,\pm}_\Pi$. 
They decay into the SM elementary particles via the intermediate composite-boson states $\Pi^0$ 
(neutral decay channel) or $\Pi^\pm$ (charged decay channel) or quarkonium states in Eqs.~(\ref{4comp}) and (\ref{4compd}). 
Suppose that at the present energy of $pp$ collisions at the LHC, composite Dirac fermions produced are non-relativistic particles, and we consider them to be approximately at rest in the CM frame of $pp$ collision. The neutral decay channels are, see Fig.~\ref{fermiond}, for the $u$-quark channel
\begin{eqnarray}
\Psi_D \rightarrow \Pi^{0} +  [\bar u_L,u_R] \rightarrow (\gamma \gamma, \gamma Z^0, W^+W^-, 
Z^0Z^0) + {\rm a~ jet},
\label{fdecay}
\end{eqnarray}
and the $d$-quark channel given by $u\rightarrow d$. 
This indicates that a composite Dirac fermion $\Psi_D$ decays
into a composite meson $\Pi^0$ and a fundamental Dirac fermion, $u$-quark $[\bar u_L,u_R]$ or $d$-quark $[\bar d_L,d_R]$. 
The latter is an ultra-relativistic quark, $E_f\approx |{\bf p}_f|$, 
ending as a jet in the final states. And the former
is a composite meson, $E_{_{\Pi^0}}=({\bf p}_{_{\Pi^0}}^2+M_{_{\Pi^0}}^2)^{1/2}$, appearing 
as an intermediate and metastable state for 
a short time $\sim 1/M_{_{\Pi^0}}$ then decays into two photons of the
energy $E_\gamma=|{\bf p}_\gamma|$ or other diboson 
channels (\ref{pidecay}) of two SM gauge bosons $G$ and $G^\prime$. 
In the CM frame of the $pp$ collision, the composite fermion is approximately at rest, and decays into a quark (jet)
and a composite boson moving apart in opposite directions (${\bf p}_{_f}=-{\bf p}_{_{\Pi^0}}$). 
The kinematic distribution of final states is 
described by the invariant mass 
${\mathcal M}_{\rm inv}\approx M_F$, 
\begin{eqnarray}
M_F&= & E_f+ E_{_{\Pi^0}}\approx |{\bf p}_{_{\Pi^0}}| +({\bf p}_{_{\Pi^0}}^2+M_{_{\Pi^0}}^2)^{1/2}, 
\nonumber\\
{\bf p}_{_f}&=&-{\bf p}_{_{\Pi^0}}=-({\bf p}_{\gamma_1}+ {\bf p}_{\gamma_2})\nonumber\\
|{\bf p}_{_{\Pi^0}}| &\approx & \Big[|{\bf p}_{\gamma_1}|^2+|{\bf p}_{\gamma_1}|^2 + 2|{\bf p}_{\gamma_1}||{\bf p}_{\gamma_2}|\cos\theta_{\gamma_1\gamma_2}\Big]^{1/2},
\label{fdecayk}
\end{eqnarray}
where $\theta_{\gamma_1\gamma_2}$ is the angle between the momenta ${\bf p}_{\gamma_1}$ and ${\bf p}_{\gamma_2}$ of two photons or other diboson states (\ref{fdecay}). 
The rate $\Gamma_{\Psi_D\rightarrow G G^\prime+{\rm jet}}$ 
of composite fermion decay (\ref{fdecay}) should be the product 
of $\Pi_0$-decay rate $\Gamma_{\Pi_{\Pi^0}\rightarrow G G^\prime}$ 
(\ref{2g}-\ref{zz}) and the rate 
$\Gamma_{\Psi_D\rightarrow \Pi^0+ {\rm a~ quark}}$ of $\Psi_D$ 
decaying into a quark and $\Pi_0$, namely 
\begin{equation}
\Gamma_{\Psi_D\rightarrow G G^\prime+{\rm a~quark}}\propto ~~\Gamma_{\Psi_D\rightarrow \Pi^0+ {\rm a~ quark}}
\times\Gamma_{\Pi_{\Pi^0}\rightarrow G G^\prime}.
\label{rfd} 
\end{equation}
This implies the ratios:
\begin{eqnarray} 
\Gamma_{\Psi_D\rightarrow \gamma Z^0+{\rm a ~quark}}/\Gamma_{\Psi_D\rightarrow \gamma\gamma 
+{\rm a ~quark}} &\propto & \Gamma_{\Pi^0\rightarrow \gamma Z^0}/\Gamma_{\Pi^0\rightarrow \gamma\gamma},\nonumber\\
\Gamma_{\Psi_D\rightarrow W^+W^-+{\rm a ~quark}}/\Gamma_{\Psi_D\rightarrow \gamma\gamma +{\rm a ~quark}}
&\propto & \Gamma_{\Pi^0\rightarrow W^+W^-}/\Gamma_{\Pi^0\rightarrow \gamma\gamma},\nonumber\\ 
\Gamma_{\Psi_D\rightarrow Z^0Z^0 +{\rm a ~quark}}/\Gamma_{\Psi_D\rightarrow \gamma\gamma +{\rm a ~quark}}&\propto  &\Gamma_{\Pi^0\rightarrow Z^0Z^0}/
\Gamma_{\Pi^0\rightarrow \gamma\gamma }.
\label{brf}
\end{eqnarray}
are similar to the ratios (\ref{br}).   
If the bosonic resonance (\ref{pidecay}) is observed and 
its invariant mass is determined, the resonance of 
composite fermion decay (\ref{fdecay})
could possibly be identified by measuring such a peculiar final-state kinematics 
(\ref{fdecayk}) of a single jet and two photons or other diboson states 
$G$ and $G^\prime$. 

\subsubsection{Decay into two gauge bosons and a lepton}\label{cfdl}

\begin{figure}[t]
\begin{center}
\includegraphics[height=2.5in, width=1.5in, angle =270 ]{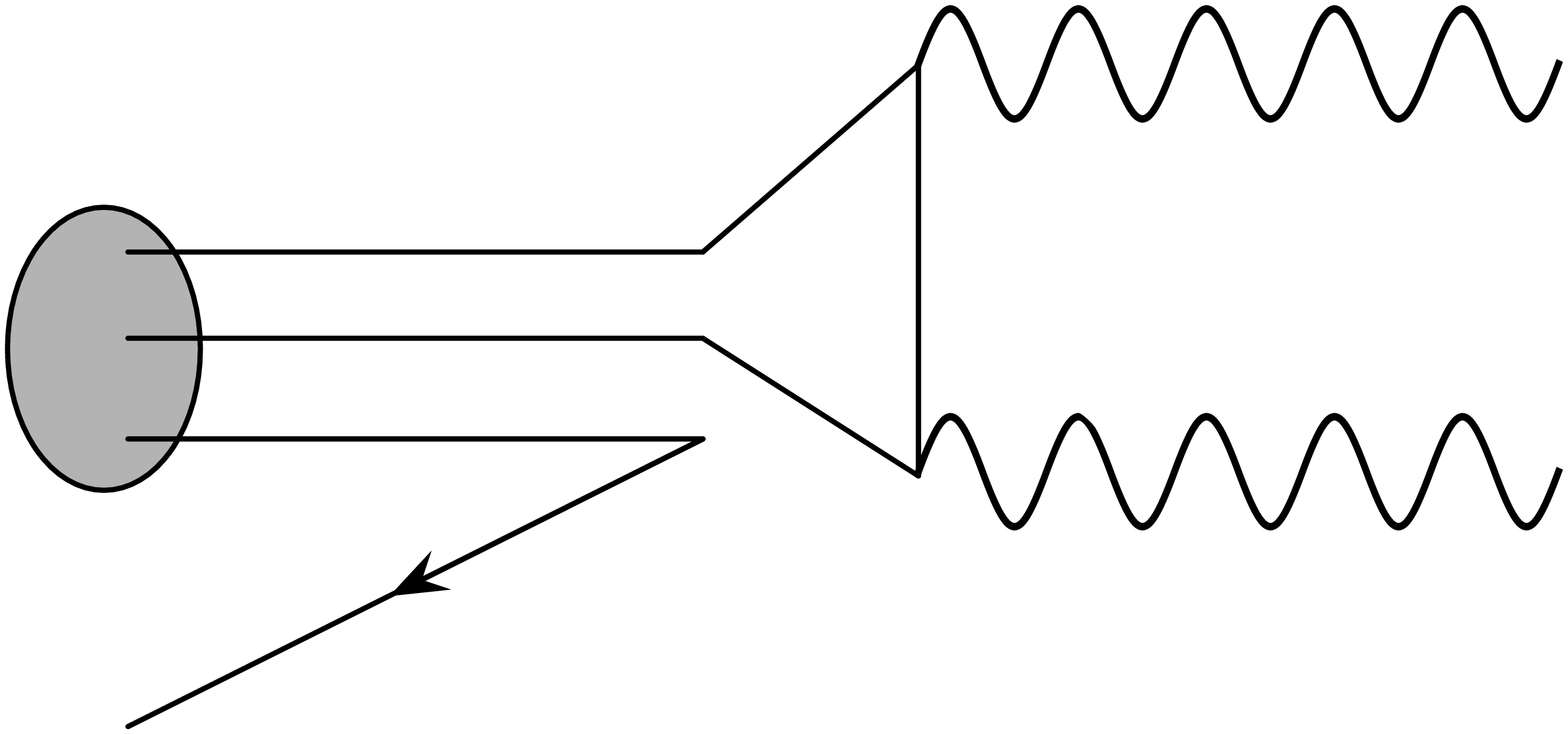}
\put(5,-7){\footnotesize $p_{\gamma_1}$}
\put(5,-62){\footnotesize $p_{\gamma_2}$}
\put(-195,-50){\footnotesize $P_F$}
\put(-90,-50){\footnotesize $\Pi^0$}
\put(-130,-95){\footnotesize $p_f$}
\caption{We show the diagram  
of a composite fermion (three parallel solid lines) decaying 
to an SM elementary quark or lepton (the solid line with an arrow) 
and a composite boson, the latter 
decays into two SM gauge bosons (wave lines), 
see Fig.~\ref{triangle}. The composite boson $\Pi^0$ is 
an intermediate state. 
$p_f$ represents the energy momentum of quark or lepton. 
$p_{\gamma_1}$ and $p_{\gamma_2}$ represent the energy momentum 
of two photons or two gauge bosons $G$ and $G^\prime$. 
The composite fermion is heavier than composite boson $\Pi^0$, 
$M_F> M_{\Pi^0}$, and the energy-momentum of composite fermion $E_F=M_F$ 
at its rest frame (${\bf P}_F=0$), as described in Eq.~(\ref{fdecayk}).}  \label{fermiond}
\end{center}
\end{figure}

Due to the four-fermion interaction of quark-lepton sector, see Eq.~(\ref{bhlql})and 
Sec.~\ref{iql}, the composite fermions (\ref{boundem}) composed by a hadronic composite boson and 
a lepton are also formed in the UV-domain. 
We generalize the analysis and discussion in the 
previous Sec.~\ref{cfdq} to the decay of composite
fermion (\ref{boundem}) composed by a hadronic composite 
boson $\Pi^0$ and a charged lepton (or a neutrino). The decay process is also represented by Fig.~\ref{fermiond}. For the $e$-lepton channel,
\begin{eqnarray}
\Psi_D \rightarrow \Pi^{0} +  [\bar e_L,e_R] \rightarrow (\gamma \gamma, \gamma Z^0, W^+W^-, 
Z^0Z^0) + {\rm an~ electron},
\label{fldecay}
\end{eqnarray}
and the $\nu_e$-lepton channel is given by $e \rightarrow \nu$. This indicates that a composite Dirac 
fermion $\Psi_D$ (\ref{boundem}) decays into a composite meson $\Pi^0$ 
and a fundamental Dirac fermion, a charged lepton 
$\ell=[\bar \ell_L,\ell_R]$ or a neutrino $\nu_{\ell}$. 
The latter is an ultra-relativistic lepton. The kinematic threshold and distribution are the same as the one (\ref{fdecay}) 
in pure hadronic channel (\ref{fdecayk}). 
However, the neutrino in the final states carries away mixing energy and momentum. 
The rate $\Gamma_{\Psi_D\rightarrow G G^\prime+{\rm a~ lepton }}$ 
of composite fermion decay (\ref{fldecay}) should be the product 
of $\Pi_0$-decay rate $\Gamma_{\Pi_{\Pi^0}\rightarrow G G^\prime}$ 
(\ref{2g}-\ref{zz}) and the rate 
$\Gamma_{\Psi_D\rightarrow \Pi^0+ {\rm a~lepton}}$ of $\Psi_D$ 
decaying into a lepton and $\Pi_0$, namely 
\begin{equation}
\Gamma_{\Psi_D\rightarrow G G^\prime+{\rm a~lepton}}\propto ~~\Gamma_{\Psi_D\rightarrow \Pi^0+ {\rm a~lepton}}
\times\Gamma_{\Pi_{\Pi^0}\rightarrow G G^\prime}.
\label{rfdl} 
\end{equation}
This implies the ratios:
\begin{eqnarray} 
\Gamma_{\Psi_D\rightarrow \gamma Z^0+{\rm a~lepton}}/\Gamma_{\Psi_D\rightarrow \gamma\gamma 
+{\rm a~lepton}} &\propto & \Gamma_{\Pi^0\rightarrow \gamma Z^0}/\Gamma_{\Pi^0\rightarrow \gamma\gamma},\nonumber\\
\Gamma_{\Psi_D\rightarrow W^+W^-+{\rm a~lepton}}/\Gamma_{\Psi_D\rightarrow \gamma\gamma +{\rm a~lepton}}
&\propto & \Gamma_{\Pi^0\rightarrow W^+W^-}/\Gamma_{\Pi^0\rightarrow \gamma\gamma},\nonumber\\ 
\Gamma_{\Psi_D\rightarrow Z^0Z^0 +{\rm a~lepton}}/\Gamma_{\Psi_D\rightarrow \gamma\gamma +{\rm a~lepton}}&\propto  &\Gamma_{\Pi^0\rightarrow Z^0Z^0}/
\Gamma_{\Pi^0\rightarrow \gamma\gamma }.
\label{brfl}
\end{eqnarray}
are similar to the ratios (\ref{br}).   

The decay processes (\ref{fdecay}) and (\ref{fldecay}) discussed in this Section show the peculiar characteristics of the effective theory of composite bosons and fermions in the UV-domain of strong four-fermion couplings. These processes are relevant to and can be checked by 
the on-going high-energy experiments in the LHC $pp$ collision.

\subsubsection{Other channels of composite fermion decay}\label{cfdc}

In addition, the decay channels of composite fermions via the intermediate meson state of charged composite bosons $\Pi^{\pm}$ (\ref{pidecay+}) read
\begin{eqnarray}
\Psi_D 
\rightarrow \Pi^{+} +  [\bar u_L,d_R],
\label{fdecayc}
\end{eqnarray}
and its charge conjugate. The Weyl fields 
$[\bar u_L,d_R]$ do not 
form a Dirac fermion of $u$- or $d$-quark, but pick up $u_R$ 
and $\bar d_L$ quarks from the vacuum, and end as 
$u$ and $d$-quark (jets) in final states. 
On the other hand, the charged composite mesons 
$\Pi^{\pm}$, see Eqs.~(\ref{4comp}) and (\ref{4compd}), 
most probably decay into two quarks 
(\ref{pidecay+}).
Therefore, the most probable final state of composite fermion 
(\ref{fdecayc}) decay 
is expected to be four jets formed by four quarks, which was 
already discussed in Refs.~\cite{xue2014,xue2015}.

These discussions can be generalized to the decay channels whose 
intermediate state is a composite quarkonium state $S^{0\pm}$, 
instead of a composite meson state $\Pi^{0\pm}$, see Eqs.~(\ref{4comp}) and (\ref{4compd}). Also, these discussion can be generalized to the second and third families of composite fermions.
  
To end this Section, we mention other possible 
channels of composite fermion decay.
Analogously to composite-fermion decay 
channel (\ref{fdecayc}), the composite fermions (\ref{boundem}) formed
the quark-lepton interaction
decay via the intermediate meson state $\Pi^{\pm}$ (\ref{pidecay+})
\begin{eqnarray}
\Psi_D 
\rightarrow \Pi^{+} +  [\bar \nu_L,\ell_R],
\label{fdecaych}
\end{eqnarray}
and its charge conjugate. The final states are a lepton pair and a jet pair.  
As for the decay of leptonic composite fermions (\ref{boundl}), final states are: (i) two gauge bosons and a lepton for the neutral 
channel via the intermediate $\Pi^0$ state; 
(ii) four leptons for the charged channel via the intermediate 
$\Pi^\pm$ state. However these decay rates are much smaller 
due to the ``missing'' of the factor $N_c^2$ and 
the smallness of the fine-structure constant $\alpha$ 
relating to the production of leptons in the LHC $pp$ collision. 

\subsection{Annihilation of two composite fermions}\label{acf}
\begin{figure}[t]
\begin{center}
\includegraphics[height=3.5in, width=1.0in, angle =270 ]{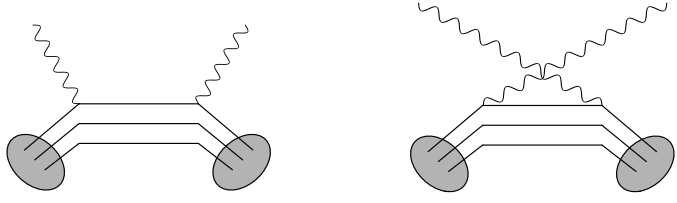}
\caption{These diagrams show the annihilation of two 
composite fermions (three solid lines) 
into two SM gauge bosons (two wave lines).} \label{fanni}
\end{center}
\end{figure}

We turn to qualitative discussions of two composite fermions 
annihilation into two SM gauge bosons $G$ and $G^\prime$, 
which should be less measurable process 
compared with a composite fermion decay, in the experimental view of producing two massive composite fermions and detecting final states.  
Analogously to the annihilation of electron and positron into 
two photons in the QED, a composite Dirac fermion and its antiparticle 
annihilates to two photons, and other massive gauge bosons, 
see Fig.~\ref{fanni},				
\begin{eqnarray}
\Psi_D+ \Psi^c_D&\rightarrow&\gamma\gamma,\gamma Z^0, W^+W^-,Z^0Z^0,\cdot\cdot\cdot
\label{anni}
\end{eqnarray}
where the composite Dirac fermion $\Psi_D$ has an electric charge $Q=(2/3,-1/3,-1)$
and mass $M_F$. 
Apart from the diphoton final state,
the final states of energetic dibosons $W^+W^-$ and $Z^0Z^0$ are two ``fat'' 
jets in opposite directions, each of them is made by 
two energetic quarks, or 
these two bosons can decay into the final states of two lepton pairs.
While in the channel $\Psi_D+ \Psi^c_D\rightarrow \gamma Z^0$, the final states are an energetic 
photon and a ``fat'' jet or an energetic photon and a lepton pair. 

Given by the mass-energy of two 
annihilating composite Dirac fermions, 
the kinematic mass-energy of final states must be larger than $2M_F$. 
This is not an invariant energy-mass representing the resonance of an 
unstable composite particle. Suppose that massive composite Dirac fermions 
produced by the LHC $pp$-collision at present energies are non-relativistic particles,
\begin{eqnarray}
E_F=(|{\bf p}_F|^2+M^2_F)^{1/2}\approx M_F + |{\bf p}_F|^2/(2M_F),
\label{nonr}
\end{eqnarray}
assuming the mass $M_F$ is large enough. 
Two SM gauge bosons $G$ and $G^\prime$ 
in the final state (\ref{anni}) are ultra-relativistic,  
the cross-section $\sigma_D $ of annihilating channel (\ref{anni}) 
can be approximately estimated by using the Dirac rate of point-like electron and positron annihilating to two photons 
with the replacement $m_e\rightarrow M_F$ and 
$\alpha\rightarrow Q^2\alpha$, 
\begin{eqnarray}
\sigma_D \sim \pi r_0^2/v, \quad r_0=Q^2\alpha/M_F, \quad  v=|{\bf p}_F|/M_F\ll 1,
\label{annicr}
\end{eqnarray}
up to the form-factors $Z^{^S}_{L,R}$ (\ref{bound}) 
of composite Dirac fermions. 
The annihilation rate $\Gamma_D$ per unit time  
\begin{eqnarray}
\Gamma_{\Psi_D\Psi^c_D\rightarrow\gamma\gamma} \approx \sigma_D v n 
\sim 2\pi r_0^2 n \sim (Q^2\alpha)^2 M_F 
\label{anninon}
\end{eqnarray}
where the number density of the produced composite fermions $n\sim M_F^3/\pi$ is assumed.
Analogously to the positronium state of electron and positron pair 
in the QED, it is possible
that the composite fermion and its antiparticle form an intermediate 
unstable Coulomb bound state, then decaying into photons or other massive 
gauge bosons. The spin singlet of the bound state decays 
into two photons (even number of photons) with the 
probability $\sim (Q^2\alpha)^5 M_F$, the spin triplet of the bound state
decays into three photons (odd number of photons) 
with the probability $\sim (Q^2\alpha)^6 M_F$.
In these discussions, we adopt two-photon final state. As for other channels of two SM gauge bosons $G$ and $G^\prime$ (\ref{anni}), 
in terms of their SM gauge couplings in the effective 
Lagrangian (\ref{comf}), we can obtain the ratios of different annihilation processes: 
\begin{eqnarray}
\Gamma_{\Psi_D\Psi^c_D\rightarrow \gamma Z^0}/\Gamma_{\Psi_D\Psi^c_D\rightarrow \gamma\gamma},\quad
\Gamma_{\Psi_D\Psi^c_D\rightarrow W^+W^-}/\Gamma_{\Psi_D\Psi^c_D\rightarrow \gamma\gamma},\quad 
\Gamma_{\Psi_D\Psi^c_D\rightarrow Z^0Z^0}/\Gamma_{\Psi_D\Psi^c_D\rightarrow \gamma\gamma},
\label{annir}
\end{eqnarray}
depending only on the SM gauge couplings $g$ and $g'$ (\ref{comf}) 
at the mass scale $M_F$, similarly to the ratios (\ref{br}). 

The following annihilation channels of two composite fermions
with different charges or zero charge are also possible, 
\begin{eqnarray}
\Psi^{-1/3}_D+ \Psi^{c2/3}_D&\rightarrow&W^-\gamma,W^- Z^0,\cdot\cdot\cdot\nonumber\\
\Psi^{0}_D+ \Psi^{0}_D&\rightarrow&W^+W^-,Z^0Z^0,\cdot\cdot\cdot
\label{anni1}
\end{eqnarray}
and their charged conjugates. 
In addition, it is known that the annihilation of electron and positron, 
through an intermediate $\gamma$-photon or $Z^0$-boson, 
produce a pair of particle and antiparticle in the SM.   
Analogously, a composite fermion and its antiparticle 
annihilates, through an intermediate
photon or $Z^0$-boson, and produce a pair of SM elementary particle and its antiparticle,  						
\begin{eqnarray}
\Psi_D+ \Psi^c_D&\rightarrow& f+ f^c,
\label{anni2}
\end{eqnarray}
where the final state is an energetic lepton pair or quark pair, 
the former is the dilepton channel and the latter is the dijet channel. Similarly to the channels (\ref{anni1}) two composite Dirac fermions
with different charges or zero charge can annihilate, through an intermediate 
charged boson $W^\pm$ or neutral boson $Z^0$, to the pair of two SM elementary fermions.

The analogous discussions on these possible processes 
can be also made for the case of the composite fermion $\Psi_S$ (singlet)
coupling to the $\gamma$ and $Z^0$, see Eq.~(\ref{comf}); as well as for the case of composite 
bosons ${\mathcal A}^i$ coupling to the $\gamma$, $Z^0$ and $W^\pm$, see Eq.~(\ref{comc}). 
In the effective Lagrangian (\ref{comf}) and (\ref{comc}), we neglect the interactions between 
the composite particles, e.g., the composite fermion and boson 
interaction of the Yukawa type. Thus we skip the discussions on two composite Dirac fermions annihilate to two composite bosons ${\mathcal A}$ and ${\mathcal A}^\dagger$. Limited by the lengthy of this article, it is impossible to discuss even qualitatively all possible processes of the effective Lagrangian (\ref{comf}) and (\ref{comc}), thus we are led to
some considerations and speculations possibly relevant to 
experiments and /or observations.  
\section
{\bf Some speculative considerations for experiments}\label{spec}

In this article, for readers' convenience, 
we first give a very brief review of 
effective four-fermion operators and effective theories 
of elementary and composite particles in IR and UV scaling domains. 
We then focus the discussions 
on the decay and annihilation channels of composite bosons and fermions 
into the final states of the SM gauge bosons, leptons and quarks, in connection with the 
searches of high-energy experiments, like the ATLAS and CMS.
The possibilities and possible criteria are provided for
the verification or falsification of such an effective theory
of massive composite particles and SM gauge couplings 
in the UV-domain of strong four-fermion couplings. 
Obviously, in addition to those processes discussed in this article, there are other possible experimentally relevant processes of 
effective four-fermion operators (\ref{art1}),   
we end this lengthy article by making some speculative considerations for experiments. 

\subsection{Some speculative considerations for LHC experiments}
In the LHC $pp$ collisions, the most probably 
channels of producing composite particles 
are the composite bosons and fermions 
(\ref{boundb}-\ref{boundm}) in the first family via 
the four-fermion operators \cite{xue2014,xue2015}
\begin{eqnarray}
G\left[(\bar\psi^{ia}_Lu_{Ra})(\bar u^b_{R}\psi^i_{Lb})
+ (\bar\psi^{ia}_Ld_{Ra})(\bar d^b_{R}\psi^i_{Lb})\right]
+G(\bar\psi^{ia}_L\nu_{R})(\bar \nu_{R}\psi^{ia}_{L}),
\label{uhlx}
\end{eqnarray}
and the channels of producing composite particles 
by other quark and lepton families have smaller rate 
because of involving small SM gauge interactions. 
The last term of Eq.~(\ref{uhlx}) represents the interaction of 
quarks and dark-matter particle $\nu_R$, which stands for the right-handed sterile neutrino of the first fermion family. 

These composite bosons and fermions could be experimentally 
verified by possibly observing their resonances and measuring their
invariant masses (${\mathcal M}_{\rm inv}$) and 
kinematic distributions of relevant final states. 
If the CM energy $\sqrt{s}$ of LHC $pp$-collisions is close to 
the masses $M_{\Pi^0}$ and $M_F$ of composite boson and fermion 
($\sqrt{s}\gtrsim M_{\Pi^0}, M_F$), 
then the invariant mass ${\mathcal M}_{\rm inv}\sim M_{\Pi^0}$ or 
${\mathcal M}_{\rm inv}\sim M_F$. The mass scales $M_{\Pi^0}$ and $M_F$
can only be determined by high-energy experiments, though we have the
theoretical relation $M_F>M_\Pi\propto \E_\xi$ of Eq.~(\ref{uscale}) and 
preliminary theoretical estimation (\ref{scales}) on the characteristic energy scale $\E_\xi\gtrsim 5 $ TeV of the UV-domain.

Suppose that the recent ATLAS and CMS preliminary 
results \cite{ATLAS2015,CMS2015} 
of diboson resonances (dijets tagged by two bosons) with invariant
masses in the energy range from 1.3 to 3.0 TeV could be further confirmed. 
These resonances are
expected to be also seen in the channels of four quark jets, 
whose invariant mass and event rate should be larger, provided that these resonances are attributed to massive composite Dirac fermions at this energy range. 
The CMS result \cite{CMSjj} of resonances with final states being two jets 
could include the event of four quark jets, two of them are geometrically 
close together to form a ``wide jet'', which should be tagged through a study 
of its substructure and flavor. Moreover, if composite Dirac fermions are formed by 
the last operator in Eq.~(\ref{uhlx}), in addition to jets in final state, 
dark-matter particles $\nu_R$ carry away missing energy-momentum 
\cite{ATLAS2015missing}.
Similar discussions are applied for the case of composite bosons.

Due to the $W^\pm$- and $Z^0$-boson couplings $g_2$ to
two constituent quarks ($u,d$) of composite fermions, in particular 
$W^\pm$-boson coupling to $SU_L(2)$ doublet 
$\psi^{ia}_L=(u^{a}_L,d^{a}_L)$,  
massive composite Dirac fermions have the following 
decay channels of final states: 
(i) dijets tagged by two highly boosted bosons WW, WZ or ZZ 
produced by high-energy constituent 
quarks ($u,d$) of composite fermions, 
together with additional quark jets; 
(ii) four quark jets formed by four high-energy constituent quarks ($u,d$) 
of a composite fermion with a peculiar kinematic 
distribution \cite{xue2014,xue2015}. It is expected that 
the former should have 
smaller rate because of the SM gauge coupling $g_2$, although we 
have not yet been able to calculate the rates of these channels. 
In these two aforementioned channels (i) and (ii), 
the final states can also be high-energy leptons, 
however the branching ratio of $W^\pm$ and $Z^0$ decaying into leptons is 
about several times smaller than that to hadrons (jets) \cite{pdg2012}.
The composite fermion can also decay 
in the channel of $W$ and Higgs (WH) bosons \cite{CMS-WH}, 
where the Higgs boson is produced by $u,d$-quarks fusing into 
a top-quark pair via a gluon, and its production rate is then 
related to the QCD coupling $\alpha_s=g^2_3/4\pi$.  
Similar discussions are applied for the case of composite bosons.
   
\subsection{Sterile neutrinos interacting with SM particles 
at high energies}
Last but not least, 
all sterile neutrinos ($\nu^i_R, \nu_R^{ic}$) and SM gauge-singlet (neutral) states of massive composite fermions, e.g., ${\bf \Psi}_D\sim [ \bar \nu^\ell_{R}, (\bar\ell^{i}_L\nu^\ell_{R})\ell_{Li}]$, can be possible candidates of warm and cold dark matter 
\cite{xue2003,xue2015}. They can couple or decay into the SM elementary particles in the following ways.
(i) SM gauge-singlet (neutral) states of composite Dirac fermions become 
unstable and decay into SM elementary particles. 
(ii) Sterile neutrinos interact with SM elementary particles
via the last term of Eq.~(\ref{uhlx}) for $\psi^i_L$ being quark or lepton 
$SU_L(2)$-doublets. 
(iii) The terms in Eqs.~(\ref{bhlxl}) and (\ref{bhlbv}) 
give the interactions
\begin{eqnarray}
G\left[ (\bar\ell^{i}_L\nu^\ell_{R})(\bar \nu^\ell_{R}\ell_{Li}) + (\bar\nu^{\ell\, c}_R\ell_{R})(\bar \ell_{R}\nu^{\ell\, c}_{R})
+(\bar\nu^{\ell\, c}_R u^{\ell}_{a,R})(\bar u^{\ell}_{a,R}\nu^{\ell c}_{R})
+(\bar\nu^{\ell\,c}_R d^\ell_{a,R})(\bar d^{\ell}_{a,R}\nu^{\ell c}_{R})\right],
\label{bhlbv'}
\end{eqnarray}
among sterile neutrinos $\nu^\ell_R,\nu_R^{\ell c}$ (dark matter) 
and SM elementary particles, where the lepton $SU_L(2)$ doublets $\ell^i_L=(\nu^\ell_L,\ell_L)$, singlets $\ell_{R}$ and the conjugate fields of sterile neutrinos 
$\nu_R^{\ell c}=i\gamma_2(\nu_R^{\ell})^*$ ($\ell=e,\mu,\tau$), and
quark fields $u^{\ell}_{a,R}=(u,c,t)_{a,R}$ 
and $d^{\ell}_{a,R}=(d,s,b)_{a,R}$. 

The four-fermion coupling $G$ in Eqs.~(\ref{uhlx}) 
and (\ref{bhlbv'}) is unique. Therefore, 
it is expected that at the same energy scale $M_F>M_\Pi\propto \E_\xi$ 
(\ref{uscale}), at which 
composite boson and fermion (\ref{boundb}-\ref{boundm}) 
appear as resonances in the LHC $pp$ collisions, 
leptonic composite boson $(\bar e_R\nu^{e c}_{R})$ or 
$(\bar \nu^{e}_Re^{}_{L})$ and composite fermion 
$[\bar\nu^{e c}_{R},(\bar e_R\nu^{e c}_{R})e_{R}]$ or 
$[\bar e^{}_{L},(\bar \nu^{e}_Re^{}_{L})\nu^e_{R}]$
should be formed by high-energy sterile neutrino inelastic collisions, 
e.g.~$\nu^e_R+\bar\nu^e_R\rightarrow e^-+e^+$ 
via the first or second interaction in Eq.~(\ref{bhlbv'}). Then these 
leptonic composite particles decay and produce electrons and positrons. 
This may account for an excess of cosmic ray electrons and positrons around TeV scale \cite{changjin2008,dark2009} in space laboratories. 
In addition, recent AMS-02 results \cite{AMS2015} 
show that at TeV scale the energy-dependent proton flux changes 
its power-law index. This implies that there would be ``excess'' 
TeV protons whose origin could be also explained by the resonance of  
composite bosons and fermions due to the 
interactions (\ref{uhlx}) and (\ref{bhlbv'}) 
of dark-matter and normal-matter particles.

We also expect that at the same energy scale 
$M_F>M_\Pi\propto \E_\xi$ (\ref{uscale}), 
the four-fermion interactions of the last term in Eq.~(\ref{uhlx}), 
the third and fourth terms in Eq.~(\ref{bhlbv'}) form
composite fermions made by $u,d$-quarks and 
sterile neutrino $\nu_R$, e.g., 
$[\bar \psi^{i}_{La},(\bar \nu_R\psi^{i}_{La})\nu_{R}]$,
$[\bar\nu^{\ell c}_{R},(\bar u_{aR}\nu^{\ell c}_{R})u_{aR}]$ or 
$[\bar u_{aL},(\bar \nu^{\ell}_Ru_{aL})\nu^\ell_{R}]$. These composite states should appear as resonances by high-energy sterile neutrinos 
inelastic collisions with nucleons (xenon) at the largest cross-section, 
then resonances decay and produce some other detectable SM particles in underground laboratories \cite{pandaX}.
Similarly, in the ICECUBE experiment \cite{icecube}, we expect the events that neutrinos change (lose) their directions (energies) by the first term of the interaction (\ref{bhlxl}) to form the resonances of composite bosons and fermions at the same energy scale 
$M_F>M_\Pi\propto \E_\xi$ (\ref{uscale}).  
In these inelastic collisions, if the accessible CM energy 
$\sqrt {s}>M_{\Pi,F}$, 
the cross section for the allowed
inelastic processes forming massive 
composite bosons and fermions will be geometrical
in magnitude, of order $\sigma_{\rm com}\sim 4\pi /M_{\Pi,F}^2$ in 
the CM frame where massive composite bosons and fermions 
are approximately at rest. 

\subsection{Further studies}
By contrast with the low-energy effective theory of SM elementary 
particles in the IR-domain of weak four-fermion coupling, 
we have to confess that the present analysis of 
the effective theory of composite particles in the UV-domain 
of strong four-fermion coupling are completely preliminary, 
due to its non-perturbative nature from the theoretical point view.
Adequate non-perturbative methods
both analytical approaches and numerical algorithms are necessary. 
This is analogous to the theoretical aspects of low-energy hadron 
physics of non-perturbative QCD. 
On the other hand, from experimental point of view, it is 
nontrivial to proceed physically sensible final-state selection 
and relevant data analysis in high-energy experiments, 
in addition to searching for high-energy collisions and accumulating 
enough data for significant analysis.  
However, without high-energy experiments, 
the characteristic energy scale $M_F>M_\Pi\propto \E_\xi$ (\ref{uscale}) 
of effective composite-particle theory 
cannot even be determined at all. 
It is worthwhile to mention that similar to the analogy between 
the Higgs mechanism and Bardeen-Cooper-Schrieffer (BCS)
superconductivity, we are studying an analogy between the effective 
theory discussed in this article and the BCS-BEC 
(Bose-Einstein condensate) crossover and unitary Fermi gas 
of strong-interacting electrons \cite{KX2017}, which is expected to 
observe in optical lattice.      

\section
{\bf Acknowledgment.}  
\hskip0.1cm The author thanks 
Prof.~Zhiqing Zhang for discussions on the LHC physics, and 
Prof.~Hagen Kleinert for discussions on the scaling domains of
the IR- and UV-stable fixed points of quantum field theories. 

\end{document}